\documentclass[]{interact}

\usepackage{epstopdf}
\usepackage[natbibapa,nodoi]{apacite}
\usepackage{hyperref}

\usepackage{tikz}
\usepackage{pgfplots,pgfplotstable}
\pgfplotsset{compat = 1.3}
\usepackage{import}
\usetikzlibrary{arrows,fit,patterns,backgrounds,positioning}
\usetikzlibrary{patterns} 
\usepgfplotslibrary{groupplots}
\usepackage{longtable}
\usepackage{multirow}

\usepackage{mathtools}

\usepackage{ marvosym}

\allowdisplaybreaks
\RequirePackage{fix-cm}
\RequirePackage{tikz}

\jyear{2022}%

\usepackage{pdflscape}
\usepackage{adjustbox}
\usepackage{placeins}
\usepackage{xcolor}

\usepackage[utf8]{inputenc}
\usepackage[T1]{fontenc}
\usepackage{amsmath,amssymb}
\PassOptionsToPackage{hyphens}{url}
\usepackage{url}
\usepackage{caption}
\usepackage{multirow}
\usepackage{setspace}
\usepackage{algorithm,algpseudocode}
\usepackage{graphicx}
\usepackage{tikz}
\usepackage{pgfplots}
\usepackage{subcaption}

\usetikzlibrary{patterns}

\usepackage{lscape}
\usepackage{longtable}
\usepackage{float}
\usepackage{breakurl}
\usepackage[official]{eurosym}
\usepackage{indentfirst}

\usepackage{afterpage,lscape}
\usepackage{rotating}

\usetikzlibrary{matrix, calc, arrows}

\usepackage{pgfplotstable} 
\usepackage{filecontents}

\pgfplotsset{
width=13cm,
compat=1.14,
}

\usepgfplotslibrary{groupplots}

\usepackage{hyperref}

\hypersetup{
  colorlinks=true,
  linkcolor=black,
  citecolor=black,
  urlcolor=black
}
\usepackage{ulem} 
\definecolor{color1}{RGB}{68, 1, 84}
\definecolor{color2}{RGB}{69, 55, 129}
\definecolor{color3}{RGB}{64, 71, 136}
\definecolor{color4}{RGB}{51, 99, 141}
\definecolor{color5}{RGB}{40, 125, 142}
\definecolor{color6}{RGB}{32, 163, 135}
\definecolor{color7}{RGB}{60, 187, 117}
\definecolor{color8}{RGB}{149, 216, 64}
\definecolor{color9}{RGB}{255, 128, 0}
\definecolor{color10}{RGB}{51, 153, 255}
\definecolor{color11}{RGB}{223, 179, 0}

\usetikzlibrary{patterns}

\begin{document}

\articletype{ORIGINAL PAPER}

\title{An efficient bundle-based approach for the share-a-ride problem}

\author{
\name{Ana Beatriz Herthel\textsuperscript{a}\thanks{CONTACT Ana Herthel Email: ana.herthel@univie.ac.at}, Richard Hartl\textsuperscript{a}, Anand Subramanian\textsuperscript{b}, Thibaut Vidal\textsuperscript{c,d}}
\affil{
\textsuperscript{a}Department of Business Decisions and Analytics, University of Vienna, Oskar-Morgenstern-Platz 1, 1090, Vienna, Austria; \textsuperscript{b}Departamento de Sistemas de Computa\c c\~ ao, Centro de Inform\' atica, Universidade Federal da Para\' iba, Rua dos Escoteiros s/n, Mangabeira, 58055-000, João Pessoa, Brazil; \textsuperscript{c}CIRRELT \& SCALE-AI Chair in Data-Driven Supply Chains, Department of Mathematical and Industrial Engineering, Polytechnique Montr\'eal, 2500 Chem. de Polytechnique, Montr\'eal, QC H3T 1J4, Canada; \textsuperscript{d}Departamento de Inform\'atica, Pontif\'icia Universidade Cat\'olica do Rio de Janeiro, Avenida Padre Leonel Franca, G\'avea, Rio de Janeiro, RJ, 22541-041}}

\maketitle

\begin{abstract}

Some of today's most significant challenges in urban environments concern individual mobility and rapid parcel delivery. With the surge of e-commerce and the ever-increasing volume of goods to be handled, new logistic solutions are in high demand. The share-a-ride problem (SARP) was proposed as one such solution, combining people and parcel transportation in taxis. This is an NP-hard problem and thus obtaining optimal solutions can be computationally costly. In this paper, we work with a variation of SARP for ride-hailing systems, which can be formulated as a multi-depot open generalised vehicle routing problem with time windows. We present and solve a mixed-integer linear programming (MILP) formulation for this problem that bundles requests together, and we compare its results to a previously proposed two-stage method. The latter solves the so-called freight insertion problem (FIP) in the second stage, for which we consider two versions, and the problem consists of inserting parcels into predefined passenger routes obtained in the first stage. We tested the methods in three sets of instances. The developed bundle-based approach outperformed both FIP versions in solution quality and in the service of parcels. Our method also compares favourably when it comes to reducing the amount of deadheading distance. 

\end{abstract}


\section{Introduction}
\label{sec:Introduction}

Growing urbanisation and the desire to reduce both the dependence on fossil energy and greenhouse gas emissions are critical factors that motivate the search for innovative solutions in urban mobility and logistics. In this context, crowd-shipping has emerged as a meaningful alternative to sharing resources and reducing carbon footprint. It relies on members of the crowd offering underused space in their vehicles to perform transportation activities.

The rapid growth of e-commerce is also shifting the way the last mile is covered for deliveries in large cities. Online stores are gaining ground over brick-and-mortar shops, given the former's inherent comfortable, easy, and rapid purchasing process. This causes an increase in the number of lightweight parcels to be delivered, which weigh less than one or two kilograms, depending on the postal service provider \citep{IPC2020}. In this context, high-capacity vehicles, such as vans and trucks, become oversized for deliveries of reduced dimensions parcels. 

The concept of sharing instead of buying
was initially expected to reduce the number of vehicles in the streets and, consequently, CO$_2$ emissions. Nevertheless, recent studies of ride-hailing systems indicate the opposite. According to \cite{Schaller2017}, \cite{TE2019}, and \cite{Anair2020}, private-hire vehicles in Europe and in the US have led to a surge in greenhouse gas emissions, and more miles travelled. A significant reason for this phenomenon is that about 42\% of their trips are associated with relocation or waiting for passenger requests and thus, are performed with an empty vehicle.


Therefore, the costly ride-hailing relocation trips can be put to better use in the transportation of parcels, guaranteeing greater efficiency in vehicle usage. This can also reduce the size of fleets dedicated to uniquely transporting packages in city centres. 

\cite{Li2014} first proposed algorithms for solving the problem of joint mobility of parcels and passengers. This problem is called the ``share-a-ride problem'' (SARP), and it arises when parcels and passengers share taxi trips.
It is an extension of the well-known dial-a-ride problem (DARP) \citep{Cordeau2003}, which is NP-hard \citep{Baugh1998}. 

The SARP is a very challenging problem as current exact methods can only obtain optimal solutions for instances with a very limited number of requests \citep{Li2014, Beirigo2018, Cavagnini2021}.  As an attempt to overcome such limitation, \cite{Li2014} proposed a two-stage solution method in which parcels are conveniently inserted into pre-determined passenger routes. This method can solve instances of larger size, but it produces suboptimal solutions because the initial passenger routes are not guaranteed to offer the best possible scheme for servicing parcels. Therefore, it is highly relevant to investigate and propose new methods capable of solving larger instances of the problem to optimality in a more efficient manner.

On the grounds of this, we have identified opportunities for generating optimal routes while simultaneously considering both parcels and passengers. The integrated optimisation approach for both types of requests is, nonetheless, a hard combinatorial problem. Hence, we devised some assumptions regarding the parcel services in the SARP, as well as a procedure that bundles requests together and cleverly groups these bundles to solve the SARP exactly, in one stage.
We further explore the capabilities of this approach by comparing it to two-stage solution methods for the SARP. 

Given the above, the main contributions of this paper are twofold:
\begin{itemize}
    \item[1.] We show that in situations in which a single passenger and a single parcel are allowed to share a vehicle, the problem can be reformulated as a multi-depot open generalised vehicle routing problem with time windows by using a bundle-based approach. 
    

    \item[2.] Through extensive computational experiments, we show that, compared to a two-stage solving technique, solving the SARP in a single stage creates more opportunities to service parcels, yields higher profits, and generates solutions with higher vehicle usage efficiency, even when some flexibility in parcel service is ceded.

\end{itemize}

The remainder of this work is organised as follows.  Section 2 reviews the relevant work related to SARP. Section 3 defines the problem and presents the necessary mathematical notation. Section 4 introduces the MILP models for the SARP. Section 5 discusses the results of our computational experiments,
and Section 6 offers concluding remarks as well as perspectives for future research.

\section{Related work}


Some recent works have presented novel ideas concerning the movement of packages via shared transportation in city streets. \cite{Li2014} proposed the SARP, an extension of the DARP, in which people and goods share trips in taxis. Each taxi can carry a single person at a time and if the driver chooses, a parcel can be picked up while transporting a person. Parcels may be collected only if the detour does not significantly affect the expected travel time of the passenger. 
\cite{Li2014} also proposed a reduced problem called the freight-insertion problem (FIP), whereby a route for servicing passengers already exists, and parcel requests can be inserted into it, if profitable. 
The SARP is solved in both dynamic and static settings, and rejections are allowed for parcels and passengers. Instances with up to 12 requests for the static SARP and up to 145 requests for the dynamic SARP and FIP were solved using a commercial MILP solver. 


\cite{Chen2017} studied the joint transportation of parcels and people, considering shopping returns. As in \cite{Li2014}, their solution involves utilising underused space in a fleet of taxis with predetermined passenger service routes. The authors impose the condition that parcels can only be collected before servicing a passenger and delivered afterward. This guarantees no detours during the passenger's trip. Furthermore, all parcels share the same final destination point, and each of them can be transported by different vehicles in their journey.

\cite{Beirigo2018} treated a variant of the SARP with shared autonomous vehicles. The resulting problem is called ``share-a-ride with parcel lockers''. For transporting people and parcels, the vehicles have distinct compartments of various sizes to accommodate goods of different dimensions, and more than one passenger can be transported at a time. The authors propose a MILP formulation for the problem and use a commercial solver to obtain solutions for instances with up to 32 requests.

The SARP addressed by \cite{Do2018} allows for a single passenger request to be serviced at a time, without detours, by considering strategies similar to the case of \cite{Chen2017}. These scenarios were studied in static and dynamic contexts in which speed windows and congestion factors were also considered.

Other variations of the SARP investigated in the literature involve the use of a mixed fleet of fossil-fuel and electric vehicles, as in \cite{Lu2022}, in which the authors propose a mathheuristic to solve the problem for a real taxi dataset of Taiwan, and also the cooperative share-a-ride introduced in \cite{Cavagnini2021}, in which the authors model the horizontal cooperation in a SARP setting for several levels of economic fairness and solve it using a commercial solver.

For a more comprehensive review of works regarding the shared trips of freight and people in different modes of transportation, as well as overviews on the state of crowd logistics and collaborative transportation, we refer the reader to the works of \cite{Mourad2019}, \cite{Sampaio2019}, \cite{Cleophas2019}, and \cite{Cavallaro2022}.

\section{Problem description}\label{sec:problem}

The SARP is the problem in which parcels and passengers share trips in taxis. In this work, we consider that this joint transportation takes place in a ride-hailing system. 
Passenger service is considered the main activity and all passenger requests must be serviced. Detours and stopovers with onboard customers are seen as detrimental to user experience \citep{Shaheen2019, Li2019, Lazarus2021}, therefore, passenger trips cannot have detours, and they do not share trips with each other. In contrast, parcel requests can be rejected if they provide no economic benefits to the system, and are outsourced to a dedicated delivery service. Hereafter, the terms passengers and customers are used interchangeably.

We define $G = (V,A)$ as a directed graph in which $V$ and $A$ represent the set of nodes and arcs, respectively. Let $n$ and $m$ refer to the number of passenger and parcel requests, respectively, and let $V$ be defined as $V = N_p  \cup N_d \cup M_p \cup M_d \cup S$, such that  $N_p = \{0, 1, \dots, n-1\}$, and $N_d = \{n, n+1, \dots, 2n-1\}$ are the set of customer requests and the set of customer pickup and drop-off nodes, respectively. Sets $M_p = \{2n, \,2n+1,\, \dots,\, 2n+m-1\}$, and $M_d = \{2n+m, \,2n+m+1,\, \dots, \,2n+2m-1\}$ contain parcel pickup and delivery nodes, respectively. Customers (parcels) picked up at $i \in N_p$ ($i \in M_p$) must be delivered to $i+n \in N_d$ ($i+m \in M_d$). 

We consider that the driver does not need to return to their original location after the last customer or parcel is serviced, as it so happens in ride-hailing services. This is commonly known as an ``open route'' and was first considered in the work of \cite{Schrage1981}. 
Nevertheless, as a modelling choice, we add ending points for vehicles to our network (``dummy depots'') that incur no travel costs and no service times. 

The fleet of homogeneous vehicles is represented by a set $K = \{0, 1, \dots, \kappa\}$. Each vehicle $k \in K$ starts its trip at a specific origin $s^k \in S$ such that $S = \{2n+2m, \,2n+2m+1,\, \dots, \,2n+2m+\kappa-1\}$ and ends it at a dummy depot $f^k \in F$, such that $F = \{2n+2m+\kappa, \,2n+2m+\kappa+1,\, \dots, \,2n+2m+2\kappa-1\}$. The vehicles also have a maximum driving time $T$. Vehicle capacity is considered through constraints imposed by the different models used in this paper.

The set of arcs, $A$, is defined by the following trips listed in Table \ref{tab:arcsdef}.
\begin{table}[!htbp]
\rm
\centering
\caption{Definition of arcs in set $A$}
\begin{tabular}{ll}
\hline
\bf{From} & \bf{To} \\ \hline
Vehicle origin  & Passenger or parcel pickup \\
Passenger pickup & Their drop-off \\
Passenger drop-off & Different passenger pickup \\
Passenger drop-off & Parcel-pickup or -delivery \\
Parcel-pickup or -delivery & Different parcel-pickup or delivery \\ \hline
\end{tabular}
\label{tab:arcsdef}
\end{table}

The arcs $(i,j) \in A$ have associated travel times ($t_{ij}$), distances ($d_{ij}$), and costs ($c_{ij}$). Values $d_{ij}$ and $t_{ij}$ are related by $d_{ij}$ = $\nu t_{ij}$, such that $\nu$ is a vehicle's average speed.

Nodes $i \in V$ are associated with a service time $r_i$, demand $q_i$, and time window $[e_i, l_i]$. The service time ($r_i$) for passengers (parcels) corresponds to the time required for boarding and leaving the vehicle (loading and unloading).
For both types of requests, $r_i > 0$ ($i \in N_p \cup N_d \cup M_p \cup M_d$), whereas $r_i = 0$ for each node $i \in S$.

The demands for parcels ($q_i$, $i \in M_p$) is equal to one and $q_i = -q_{i+m}$ ($i \in M_p$). For customers, $q_i = 3$ and $q_i = -q_{i+n}$ ($i \in N_p$). The demands for $i \in S$ have null values.

We consider passenger time windows as time points such that $e_i = l_i$, and feasibility of service is guaranteed by defining $e_{i+n} = l_{i+n} = e_i + r_i + t_{i,i+n}$.
The remaining nodes have time windows set to the complete planning horizon.

The objective of the SARP is to maximise profit by simultaneously transporting passengers and parcels. Further necessary parameter definition is presented in Table~\ref{tab:parampresent}.

\begin{table}[!htbp]
\rm
\centering
\caption{Parameters for calculating profit}
\begin{tabular}{cl}
\hline
\bf{Parameter} & \bf{Description} \\ \hline
$\gamma_1$ & Initial fare charged for passenger transportation \\
$\mu_1$& Fare per km charged for passenger transportation \\
$\gamma_2$ & Initial fare charged for parcel transportation \\
$\mu_2$ & Fare per km charged for parcel transportation \\
$\mu_{3}$ & Average driving costs per km \\ 
$\phi_i$ & Revenue for transporting  passenger $i\in N_p$ \\
$\theta_i$ & Revenue for transporting  parcel $i\in M_p$\\ \hline
\end{tabular}
\label{tab:parampresent}
\end{table}

The travel costs are estimated to be proportional to the distance: $c_{ij} = \mu_3 d_{ij}$. We assume fuel, insurance, and maintenance costs are included in $\mu_{3}$.


\section{Methodology}

This section describes the two-stage and one-stage approaches implemented in this work. For the former, we discuss changes to the existing freight insertion problem, while for the latter, we present our newly proposed bundle-based MILP formulation.

\subsection{Two-stage approach and the freight insertion problem}

We consider the two-stage approach proposed in \cite{Li2014} for the SARP, which starts with a feasible solution to the problem, exclusively considering the service to customers. 
From this initial solution, the authors define that a single parcel service node (either pickup or delivery) can be inserted between any two customer service ones. It is worth noting that the available parcel insertion positions, in this problem, do not include the possibility of servicing these requests before the first customer is picked up or after the last customer is dropped off. The problem of determining the optimal insertion of the parcels given a set of initial passenger-only routes is known as FIP.

The FIP in \cite{Li2014} allows for the possibility of detours during customer trips, which are subject to a discount factor in the objective function. Nevertheless, the authors restrict the duration of these detours to twice the direct travel time between pickup and drop-off locations for a particular customer and the capacity of vehicles to five.

As in the original FIP, we also assume that at most one parcel node (either pickup or delivery) can be visited before a customer is picked up or after they are dropped off. However, we extend parcel service opportunities to allow for their insertion before the first customer is picked up and after the last customer's trip is ended. This favours parcel service even in short initial routes, with few customer requests. As mentioned previously, in our version of SARP, we do not allow for passenger detours. Furthermore, the initial routes, in our case, are obtained by solving a simple MILP model for customer requests only. We refer to \cite{Li2014} for the FIP model and its related notation.

\subsection{One-stage approach: bundle formulation}

In order to solve the SARP in a single stage and in reasonable computational times, we took inspiration from FIP in some of its assumptions, as well as proposing others to devise an improved solution approach.

First, we limit the vehicle capacity such that at most one parcel can be carried at a time. However, unlike FIP, we determine that a parcel picked up must be delivered immediately after exactly one passenger's trip ends. This ensures that there will be space in the vehicle for passengers' belongings and the possibility of delivering more time-sensitive parcels.

\begin{figure}[!htbp]
    \centering
        \includegraphics[scale=0.3]{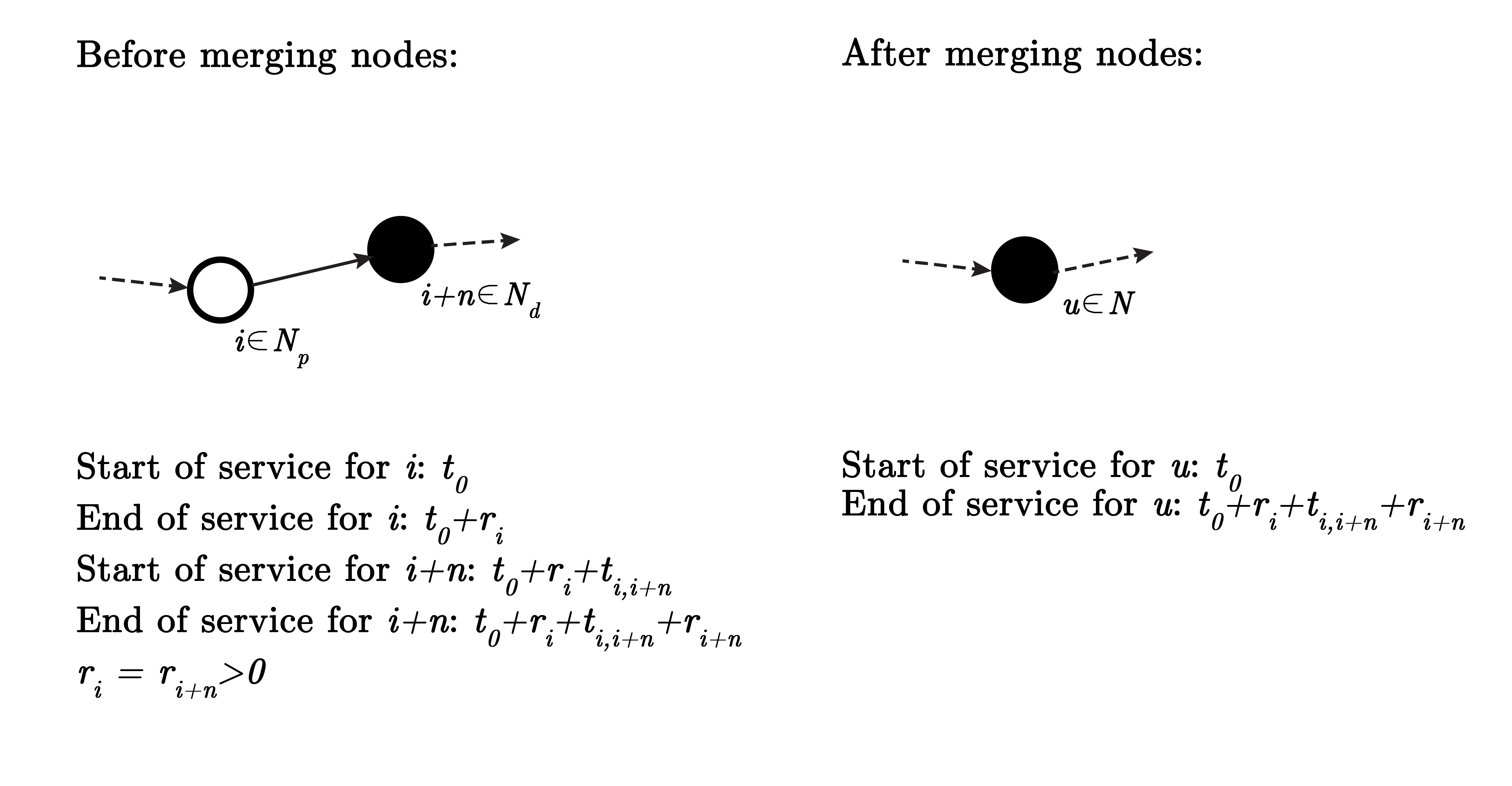}
    \caption{Example of customer node merging}
    \label{fig:mergednodes}
\end{figure}

Additionally, since passengers do not have detours, a passenger pickup is always followed by their drop-off.
Hence, for modelling purposes, these two activities are represented as a single node and sets $N_p$ and $N_d$ are merged into set $N = \{0, 1, \dots, n-1\}$, such that each node pair $(i, i+n)$, $i \in N_p, i+n \in N_d$, becomes a node $u \in N$, as shown in Figure~\ref{fig:mergednodes}.

We further observe that a vehicle leaving its origin can either visit a customer or a parcel-pickup location next. In the latter case, a vehicle visiting a node $v \in M_p$ follows up by visiting $u \in N$ and, later, $v+m \in M_d$. This is true at every visit of a node $v \in M_p$. After this point, the visiting possibilities of the vehicle are restored to the initial options (customer node or parcel-pickup location). The same initial options are also available to follow up the servicing of any customer node, provided that the vehicle is not carrying any parcels at the time.
 
Considering these observations, we arrange the available requests to form ``service bundles''. These bundles can contain either a single customer request $u \in N$ or a contiguous subsequence of requests $(v,u,v+m)$ such that $v \in M_p$, $u \in N$, and $v+m \in M_d$.

Inspired by the generalised vehicle routing problem (GVRP) defined by \cite{Ghiani2000}, we further separate the formed bundles into groups. The number of customer requests determines the number of groups, which are comprised of all bundles containing a particular customer request.
Therefore, we can guarantee that all customers are serviced by selecting a single bundle from each group.

\begin{figure}[!htbp]
    \begin{center}
        \includegraphics[scale=0.35]{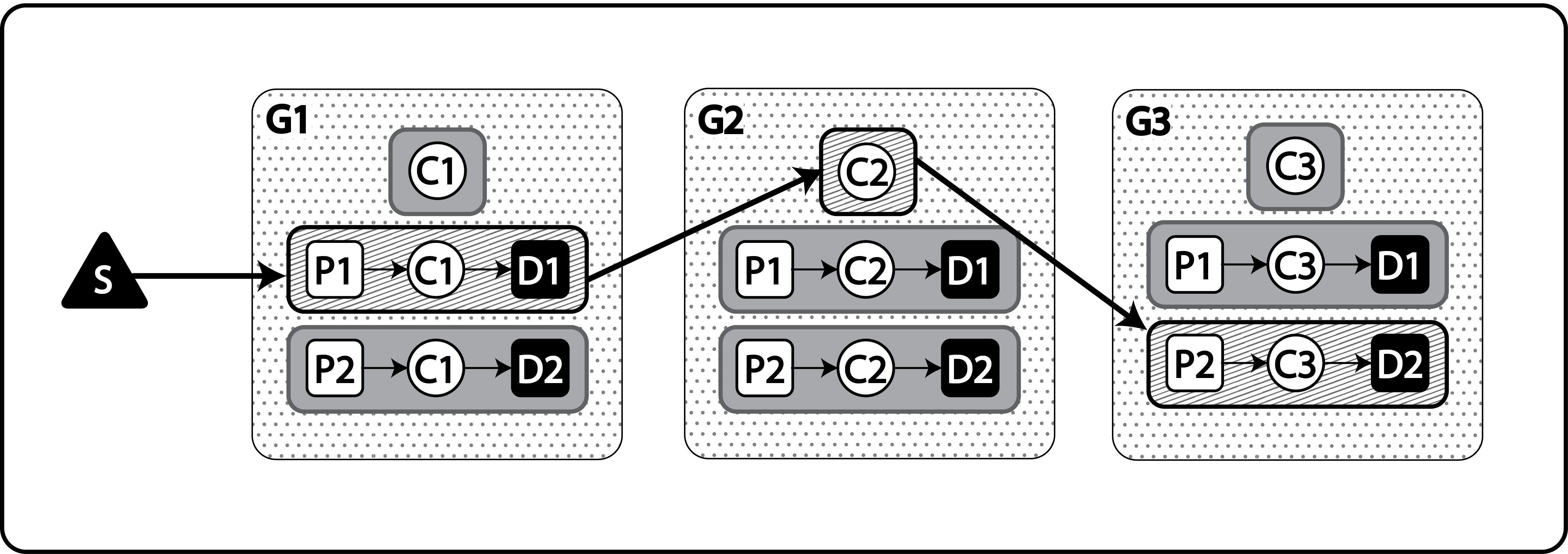}
\includegraphics[scale=0.4]{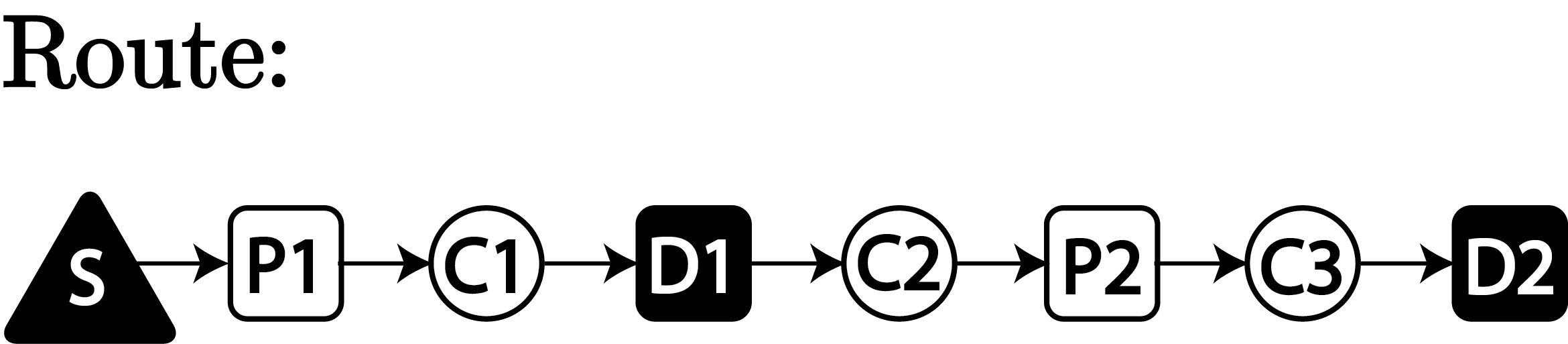}
    \end{center}
\hspace*{1.85cm}
\includegraphics[scale=0.4]{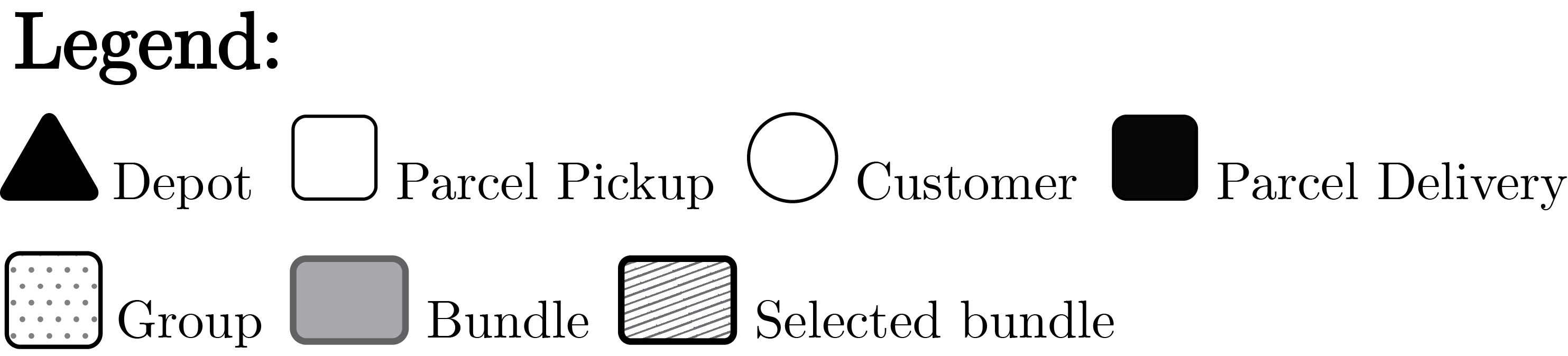}
    \caption{Example of bundle grouping and solution construction}
    \label{fig:bundleex}
\end{figure}

Figure~\ref{fig:bundleex} shows an example of grouping bundles and constructing solutions with the new arrangement.

Hence, we define a graph $\bar{G} = (B,A_B)$ such that each node corresponds to one of the possible bundles. The set of bundles is represented by $B = \{0,\, 1, \,\dots, \,n+nm-1\}$ and $A_B=\{(i,j) : i \in B\cup S\cup F,\, j \in B\cup S\cup F,\, i \neq j\}$ defines the set of arcs between bundles, origins, and dummy depots for vehicles. Furthermore, we define $B_u\subset B$ and $B_v\subset B$ as the sets of bundles respectively servicing passenger $u \in N$ and parcel $v \in M_p$.

A vehicle $k \in K$ has to start and end its trip at its specific origin and ending locations. These points are not bundled with any other requests in this formulation. 

The revenues $\beta_i$ generated by a bundle $i \in B$ are calculated by adding individual revenues for the bundled requests and subtracting the travelling costs between them. 

Each bundle also has a service time $\Delta_i$ and a time window $[\bar{e}_i, \bar{l}_i]$ that needs to be satisfied. Since passengers are bound to be serviced at specific time points, no bundle can be serviced later than $\bar{e}_i$, so $\bar{e}_i = \bar{l}_i$ for $i \in B$. The time  $\bar{e}_i$  to start servicing  bundle $i \in B$ is obtained from the time point of service for the passenger associated with bundle $i$ (${e}_u, u \in N$). If no request requires service before the passenger, the bundle must be serviced at ${e}_u$. Otherwise, $\bar{e}_i = e_{u} - t_{vu} - r_{v}$ such that $u \in N$, $v \in M_p$, and $i \in B$ and $u$ and $v$ are requests present in $i$. The service time $\Delta_i$, $i \in B$ is calculated as the sum of the service times of all requests present in the bundle and the travel times between them.

Each arc $(i,j)\in A_B$ is characterised by costs $\bar{c}_{ij} > 0$ and travel times $\bar{t}_{ij} > 0$. For cases where an arc leaves a bundle to proceed to a dummy depot, both values are set to zero.
The arcs $(i,j) \in A_B$ represent trips from $i$ to $j$ ($i,j \in B\cup S\cup F$) and  exist in  the following cases:
\begin{itemize}
    \item[1.] from a vehicle origin to a bundle;
    \item[2.] between any two bundles that do not contain the same passenger request;
    \item[3.] from a bundle to a dummy depot.
\end{itemize}

Note that arcs are removed from the network if $\bar{e}_i < 0$, $\bar{e}_i + \Delta_{i} > \tau$ ($i \in B$), or both. 

The  graph described is used to develop the ``Bundle formulation'' ($BF$). The decision variables $z_{ij}^k$ in this formulation are unity if arc $(i,j) \in A_B$ is traversed by vehicle $k \in K$ and zero otherwise. The resulting model is 

\begin{align}
 \max\sum\limits_{i \in B}\sum\limits_{\substack{j\in B}} \sum\limits_{k\in K} \beta_{i}z_{ij}^k - \sum\limits_{\substack{(i,j)\in A_B}}\sum\limits_{k\in K} \bar{c}_{ij}z_{ij}^{k} \label{eq:bfoio}
\end{align}
subject to
\begin{align}
    &\sum\limits_{k\in K}\sum\limits_{i\in B_u}\sum\limits_{\substack{j\in B}}z_{ij}^k = \sum\limits_{k\in K}\sum\limits_{\substack{i\in B}}\sum\limits_{j\in B_u}z_{ij}^k = 1, & & u \in N,\label{eq:br1}\\
    &\sum\limits_{j\in B}z_{(s_k,j)}^k = \sum\limits_{i\in B}z_{(i,{f}^k)}^k = 1, & & k \in K,\label{eq:br2}\\
    &\sum\limits_{i\in B_{v}}\sum\limits_{\substack{j\in B}}\sum\limits_{k\in K}  z_{ij}^k \leq 1, & & v \in P,\label{eq:br4}\\
    &\sum\limits_{\substack{j\in B}}z_{ij}^k = \sum\limits_{\substack{j\in B}}z_{ji}^k ,& & i\in B,\quad k \in K,\label{eq:br5}\\
    &\sum\limits_{i \in B} (\bar{e}_i + \Delta_i)z_{(i,{f}^k)}^k - \sum\limits_{\substack{j\in B}}(\bar{e}_j - \bar{t}_{(s_k,j)})z_{(s_k,j)}^k \leq T, && k \in K, \label{eq:br6}\\
    &z^k_{ij} \in \{0,1\}, & &  (i,j)\in A_B, \quad k \in K. \label{eq:br7}
\end{align}

 Objective \eqref{eq:bfoio} maximises profit for visiting bundles. Constraints \eqref{eq:br1} establish that, for each passenger request, a single arc leaves and enters the group of bundles containing it. Constraints \eqref{eq:br2} determine each route's origin and end points. Constraints \eqref{eq:br4} ensure that no parcel is serviced more than once, and Constraints \eqref{eq:br5} guarantee flow conservation between bundles. Constraints \eqref{eq:br6} ensure that the maximum driving time is not exceeded, whereas Constraints \eqref{eq:br7} determine the variables domain.
 
\section{Computational experiments}\label{sec:compexp}

We implemented the models using C++ and g++ 5.4.0, compiled with flag -O3. We conducted all computational experiments on a single thread of a server equipped with an Intel Xeon 2.0 GHz processor and 128 GB of RAM, running Ubuntu Linux 16.04. We used IBM CPLEX 12.7 to solve the models and established a limit of two hours for each run. 

The goal of our experiments is to provide information for further assessment of cost, service, computational and environmental impacts regarding the different decision strategies, namely, FIP and $BF$, proposed for the considered versions of SARP. 

Our evaluation starts by focusing on two critical aspects of the system: level of service and cost. Since all passengers have to be serviced, the former is measured in terms of parcels serviced, while the latter accounts for profit obtained in the shared transportation service (objective function value).

Furthermore, we test the scalability of the FIP and $BF$, observing how instance size variation impacts each method's solving times.
Lastly, we consider
the environmental aspects of the problem by gauging how the parcel service affects the use of resources; that is, how much distance a vehicle travels empty, which is referred to as ``deadheading'' henceforth \citep{Lazarus2021, Henao2019}.

\subsection{Calibration of the cost parameters}

We gathered Uber data regarding trip price estimates (\url{https://www.uber.com/global/en/price-estimate/}) from cities with a significant number of daily ride requests and parcel deliveries. This information was used to calculate parameters $\gamma_1$ and $\mu_1$, as presented in Table \ref{tab:params1}. All monetary amounts were converted to US dollars, and the fare per mile was converted to fare per kilometre.


\begin{table}[!htbp]
\footnotesize
  \centering
  \caption{Definition of $\gamma_1$ and $\mu_1$}
    \begin{tabular}{lcc}
    \hline
    \bf{City} & \bf{$\gamma_1$ (US\$)} & \bf{$\mu_1$ (US\$)} \\\hline
    {New York} & 0.00 & 0.92 \\
    {Los Angeles} & 0.00  & 0.50 \\
    {San Francisco} & 2.20 & 0.57 \\
    {London} & 13.10 & 1.31 \\
    {Berlin} & 2.36  & 1.49 \\
    {Paris} & 1.42 & 1.24 \\
    {Vienna} & 11.80 & 0.77 \\
    {Rome} & 5.90 & 1.31 \\
    {São Paulo} & 0.34  & 0.24 \\
    {Rio de Janeiro} & 0.36  & 0.25 \\
    {Tokyo} & 0.98  & 3.13 \\
    {Sidney} & 1.80  & 1.04 \\
    {Toronto} & 1.90 & 0.62 \\\hline
      \bf{Avg.}  & \bf{3.24} & \bf{1.03} \\\hline
    \end{tabular}%
  \label{tab:params1}%
\end{table}%

We also collected data on costs of fuel, insurance, and maintenance from large countries and the European Union to compute $\mu_3$.
The values were converted to US dollars, and a simple average provided the final value of $\mu_3$ as presented in Table~\ref{tab:parammu3}.

\begin{table}[!htbp]
\footnotesize
  \centering
  \caption{Definition of $\mu_3$}
    \begin{tabular}{lc}
    \hline
    \bf{Database} & \bf{Costs per km (US\$)} \\\hline
    United States  & 0.36 \\
    European Union   & 0.48 \\
    Brazil & 0.41 \\
    Canada & 0.58 \\
    \hline
    \textbf{Avg.} & \textbf{0.46} \\\hline
    \end{tabular}%
  \label{tab:parammu3}%
\end{table}%

To estimate $\gamma_2$ and $\mu_2$, we consider that e-commerce consumers hope to pay as little as possible for transporting parcels. At the same time, the integration of parcel-delivery services should be profitable for ride-hailing companies. To this end, we use $\gamma_2 = 2.74$ and $\mu_2 = 0.83$, which are estimated as a reduction of approximately 15\% and 20\% in values for $\gamma_2$ and $\mu_2$, respectively, compared to their counterparts in passenger service. The definition of these values ensures passenger transportation remains the most profitable service in the problem while also guaranteeing a meaningful impact on the objective function value and parcel service profitability ($\mu_2 > \mu_3$).

From data provided by the Uber Movement initiative \citep{ubermov}, we  also obtained information regarding average speed and average trip duration for Uber requests, where available. 
We selected October 2019 as our data source because this month offers the most recent complete set for which customer behaviour is not affected by the Covid-19 pandemic or holiday season. Table~\ref{tab:param4} lists the values and averages.

\begin{table}[!htbp]
\centering
\setlength{\tabcolsep}{0.1cm}
\footnotesize
\caption{Defining $\nu$ and distance range}
\begin{tabular}{lccccccccc}
    \hline
     \multirow{4}{*}{\bf City} &  {\bf Speed} & &  {\bf Travel} & & \multicolumn{2}{c}{\bf Distance} & &  \multicolumn{2}{c}{\bf Distance} \\
     & {\bf (km/h)} & &  {\bf time (h)} & & \multicolumn{2}{c}{\bf (km)} & &  \multicolumn{2}{c}{\bf range (km)} \\\cline{2-2} \cline{4-4}\cline{6-7}\cline{9-10}
    & {\bf Avg.} & & \bf Avg. & & \bf Avg. & \bf Std. Dev. & & {\bf Min} & {\bf Max}\\\hline
    São Paulo & 29.451 & & 0.597 & & 17.590 & 1.198 & & 16.392 & 18.788 \\
    Seattle & 47.826&  & 0.305 & & 14.598 & 0.832 & & 13.766 & 15.430 \\
    London & 34.597 & & 0.485 & & 16.778 & 0.909 & & 15.869 & 17.686 \\
    Madrid & 45.296 & & 0.263 & & 11.935 & 1.091 & & 10.844 & 13.027 \\
    San Francisco & 47.544 & & 0.438 & & 20.825 & 1.126 & & 19.699 & 21.951 \\\hline
    \textbf{Avg.} & \textbf{40.943} &  &  \textbf{0.418} & & \textbf{16.345} &  \textbf{1.031} & & \textbf{15.314} & \textbf{17.376} \\\hline
    \end{tabular}%
  \label{tab:param4}%
\end{table}%

\subsection{Test instances}\label{sec:Inst}

We worked with five datasets divided into two classes, based on their sizes. The first set, $GH_1$ was adapted from the Travelling Salesman Problem Pickup and Delivery Test Instance Library (TSPPDLIB) of \cite{grubhub} (\url{https://github.com/grubhub/tsppdlib}), based on a meal-delivery application from Grubhub. This library includes ten instances for each number of requests, which spans from two to fifteen pickup and delivery pairs. We adapted instances from this library with a minimum of eleven requests, dividing them into customers and parcels.

We generated points in time for passenger requests and considered our time horizon to be 24 hours. 
Further adaptation was necessary to scale the TSPPDLIB distances by means of a multiplicative factor to obtain average distances of a magnitude similar to those from the Uber data.

The second and third sets, $SF_1$ and $SF_2$ are derived from instances of sizes 270 and 300, presented in \cite{Li2016a}. We selected a number of passenger and parcel requests to form smaller instances of our own. For $SF_1$ and $SF_2$, we respectively generated instances of sizes ranging from 10 to 15 and from 20 to 60. Five different instances were created for each number of requests.

The last two sets, $MD_1$ and $MD_2$ are introduced in this work. We generated points in a grid with distances corresponding to the ranges calculated from Uber data. We paired these points as pickup and destination using a method similar to that proposed by \cite{Renaud2000}.
Six multiple depot instances are generated for each combination of number of passenger and parcel requests.

The number of customers and parcels chosen for sets of class $1$ ($GH_1$, $SF_1$, $MD_1$) varies from 5 to 10 for each type of request, with a total ranging from 10 to 15 requests.
For sets of class $2$ ($MD_2$ and $SF_2$) the number of requests for customers varies from the values in the set $R = \{5, 10, 15, 20, 30\}$, and parcels, according to $R \setminus \{5\}$.

We chose datasets derivative from multiple sources as a way to test the robustness of our proposed solution method.

\subsection{Service and cost impacts}

We analyze the performance of the $BF$ by comparing it to two variations of the FIP formulation, namely, $FIP_{sg}$ and $FIP_{mt}$. In $FIP_{sg}$ 
we define the maximum capacity for parcel carrying as one, similar to the $BF$, while $FIP_{mt}$ 
has no upper limit to parcel carrying capacity.
Both FIP variations have no restrictions regarding when parcels are delivered.

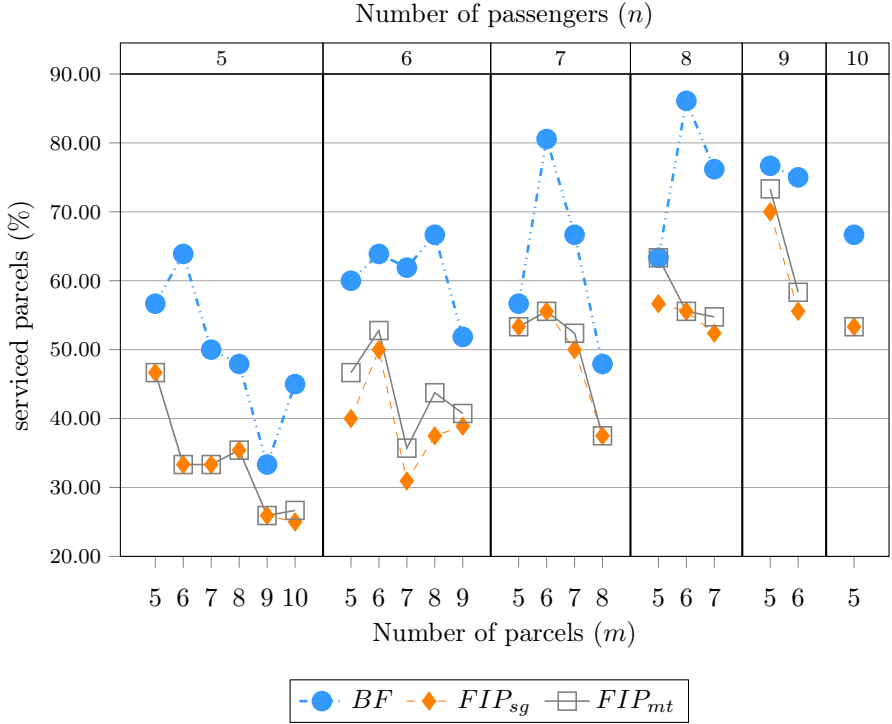
\begin{figure}[!htbp]
    \centering
    \footnotesize

    \pgfplotstableread[col sep=&, header=true]{
description     &       $BF$ &      $FIP_{sg}$ & $FIP_{mt}$
Sarp-5-5 & 56.67 & 46.67 & 46.67  
Sarp-5-6 & 63.89 & 33.33 & 33.33  
Sarp-5-7 & 50.00 & 33.33 & 33.33  
Sarp-5-8 & 47.92 & 35.42 & 35.42  
Sarp-5-9 & 33.33 & 25.93 & 25.93  
Sarp-5-10 & 45.00 & 25.00 & 26.67  
          & nan & nan & nan
Sarp-6-5 & 60.00 & 40.00 & 46.67  
Sarp-6-6 & 63.89 & 50.00 & 52.78  
Sarp-6-7 & 61.90 & 30.95 & 35.71  
Sarp-6-8 & 66.67 & 37.50 & 43.75  
Sarp-6-9 & 51.85 & 38.89 & 40.74
 & nan & nan & nan
Sarp-7-5 & 56.67 & 53.33 & 53.33  
Sarp-7-6 & 80.56 & 55.56 & 55.56  
Sarp-7-7 & 66.67 & 50.00 & 52.38  
Sarp-7-8 & 47.92 & 37.50 & 37.50 
 & nan & nan & nan
Sarp-8-5 & 63.33 & 56.67 & 63.33  
Sarp-8-6 & 86.11 & 55.56 & 55.56  
Sarp-8-7 & 76.19 & 52.38 & 54.76  
 & nan & nan & nan
Sarp-9-5 & 76.67 & 70.00 & 73.33  
Sarp-9-6 & 75.00 & 55.56 & 58.33
 & nan & nan & nan
Sarp-10-5 & 66.67 & 53.33 & 53.33  
}\datatableentry
    \begin{tikzpicture}
        \begin{axis}[
            height=8cm, width=11.8cm,
            ymin=20, ymax=90,
            ytick={20,30,...,90},
            grid=both,
            major grid style={line width=.1pt,draw=gray!70},
            xmajorgrids=false,
            tick align=outside,
            title style={at={(0.5,1)},anchor=south,yshift=8 },
            title={\small Number of passengers ($n$)},
            tickpos=left,
            xtick=data,
            xticklabels ={5, 6, 7, 8, 9, 10, 5, 6, 7, 8, 9, 5, 6, 7, 8, 5, 6, 7, 5, 6, 5},
            enlarge x limits = 0.05,
            yticklabel style = { 
                /pgf/number format/fixed zerofill,
                /pgf/number format/precision=2,
            },
            legend columns=5,
            legend columns=7,
            legend style={
                at={(0.48,-0.3)},
                anchor=center,
                draw=black,
                legend cell align={center},
                /tikz/every even column/.append style={font=\small, column sep=0.1cm}
            },
            ylabel={\small serviced parcels (\%)},xlabel={\small Number of parcels ($m$)},
            every x tick label/.append style={font=\small,yshift = -5 pt},
            extra description/.code={
            \draw (axis cs: -1.25,90) rectangle (axis cs: 6,94.5) node[pos=.5] {5};
            \draw (axis cs: 6,90) rectangle (axis cs: 12,94.5) node[pos=.5] {6};
            \draw (axis cs: 12,90) rectangle (axis cs: 17,94.5) node[pos=.5] {7};
            \draw (axis cs: 17,90) rectangle (axis cs: 21,94.5) node[pos=.5] {8};
            \draw (axis cs: 21,90) rectangle (axis cs: 24,94.5) node[pos=.5] {9};
            \draw (axis cs: 24,90) rectangle (axis cs: 26.25,94.5) node[pos=.5] {10};
            }
        ]
        \addplot [dashdotdotted, line width=1pt, color=color10,mark=*, unbounded coords=jump, mark options={scale=1.7,solid}] table [y=$BF$, x expr=\coordindex] {\datatableentry};
        \addplot [dashed,color=color9,mark=diamond*, unbounded coords=jump, mark options={scale=1.7,solid}] table [y=$FIP_{sg}$, x expr=\coordindex] {\datatableentry};
        \addplot [semithick,color=gray,mark=square, unbounded coords=jump, mark options={scale=1.7,solid}] table [y=$FIP_{mt}$, x expr=\coordindex] {\datatableentry};

        \addplot[thick, black, forget plot] coordinates {(6,20)(6,90)};
        \addplot[thick, black, forget plot] coordinates {(12,20)(12,90)};
        \addplot[thick, black, forget plot] coordinates {(17,20)(17,90)};        \addplot[thick, black, forget plot] coordinates {(21,20)(21,90)};
        \addplot[thick, black, forget plot] coordinates {(24,20)(24,90)};
        
    \legend{$BF$,$FIP_{sg}$,$FIP_{mt}$}
    \end{axis}
    \end{tikzpicture}
    \caption{Amount of serviced parcels (\%) for $MD_1$ dataset}
    \label{fig:servedsmd}
\end{figure}

\begin{figure}

    \centering
    \footnotesize

    \pgfplotstableread[col sep=&, header=true]{
description     &       $BF$ &      $FIP_{sg}$ & $FIP_{mt}$

Sarp-5-5 & 66.00 & 42.00 & 44.00  
Sarp-5-6 & 55.00 & 31.67 & 33.33  
Sarp-5-7 & 54.29 & 40.00 & 42.86  
Sarp-5-8 & 50.00 & 35.00 & 35.00  
Sarp-5-9 & 46.67 & 26.67 & 27.78  
Sarp-5-10 & 45.00 & 26.00 & 26.00  
          & nan & nan & nan
Sarp-6-5 & 60.00 & 40.00 & 42.00  
Sarp-6-6 & 71.67 & 50.00 & 53.33  
Sarp-6-7 & 58.57 & 40.00 & 44.29  
Sarp-6-8 & 55.00 & 38.75 & 38.75  
Sarp-6-9 & 54.44 & 37.78 & 38.89  
          & nan & nan & nan
Sarp-7-5 & 82.00 & 66.00 & 72.00  
Sarp-7-6 & 68.33 & 48.33 & 48.33  
Sarp-7-7 & 71.43 & 47.14 & 47.14  
Sarp-7-8 & 68.75 & 38.75 & 41.25  
          & nan & nan & nan
Sarp-8-5 & 80.00 & 62.00 & 62.00  
Sarp-8-6 & 78.33 & 46.67 & 50.00  
Sarp-8-7 & 71.43 & 52.86 & 54.29  
          & nan & nan & nan
Sarp-9-5 & 84.00 & 62.00 & 66.00  
Sarp-9-6 & 90.00 & 68.33 & 70.00  
          & nan & nan & nan
Sarp-10-5 & 94.00 & 62.00 & 66.00  
}\datatableentry
    \begin{tikzpicture}
        \begin{axis}[
            height=8cm, width=11.8cm,
            ymin=20, ymax=100,
            ytick={20,30,...,100},
            grid=both,
            major grid style={line width=.1pt,draw=gray!70},
            xmajorgrids=false,
            tick align=outside,
            title style={at={(0.5,1)},anchor=south,yshift=8 },
            title={\small Number of passengers ($n$)},
            tickpos=left,
            xtick=data,
            xticklabels ={5, 6, 7, 8, 9, 10, 5, 6, 7, 8, 9, 5, 6, 7, 8, 5, 6, 7, 5, 6, 5},
            enlarge x limits = 0.05,
            yticklabel style = { 
                /pgf/number format/fixed zerofill,
                /pgf/number format/precision=2,
            },
            legend columns=5,
            legend columns=7,
            legend style={
                at={(0.48,-0.3)},
                anchor=center,
                draw=black,
                legend cell align={center},
                /tikz/every even column/.append style={font=\small, column sep=0.1cm}
            },
            ylabel={\small serviced parcels (\%)},xlabel={\small Number of parcels ($m$)},
            every x tick label/.append style={font=\small,yshift = -5 pt},
            extra description/.code={
            \draw (axis cs: -1.25,100) rectangle (axis cs: 6,104.5) node[pos=.5] {5};
            \draw (axis cs: 6,100) rectangle (axis cs: 12,104.5) node[pos=.5] {6};
            \draw (axis cs: 12,100) rectangle (axis cs: 17,104.5) node[pos=.5] {7};
            \draw (axis cs: 17,100) rectangle (axis cs: 21,104.5) node[pos=.5] {8};
            \draw (axis cs: 21,100) rectangle (axis cs: 24,104.5) node[pos=.5] {9};
            \draw (axis cs: 24,100) rectangle (axis cs: 26.25,104.5) node[pos=.5] {10};
            }
        ]
        \addplot [dashdotdotted, line width=1pt, color=color10,mark=*, unbounded coords=jump, mark options={scale=1.7,solid}] table [y=$BF$, x expr=\coordindex] {\datatableentry};
        \addplot [dashed,color=color9,mark=diamond*, unbounded coords=jump, mark options={scale=1.7,solid}] table [y=$FIP_{sg}$, x expr=\coordindex] {\datatableentry};
        \addplot [semithick,color=gray,mark=square, unbounded coords=jump, mark options={scale=1.7,solid}] table [y=$FIP_{mt}$, x expr=\coordindex] {\datatableentry};

        \addplot[thick, black, forget plot] coordinates {(6,20)(6,100)};
        \addplot[thick, black, forget plot] coordinates {(12,20)(12,100)};
        \addplot[thick, black, forget plot] coordinates {(17,20)(17,100)};        \addplot[thick, black, forget plot] coordinates {(21,20)(21,100)};
        \addplot[thick, black, forget plot] coordinates {(24,20)(24,100)};
        
    \legend{$BF$,$FIP_{sg}$,$FIP_{mt}$}
    \end{axis}
    \end{tikzpicture}
    \caption{Amount of serviced parcels (\%) for $GH_1$ dataset}
    \label{fig:servedsgh}
\end{figure}
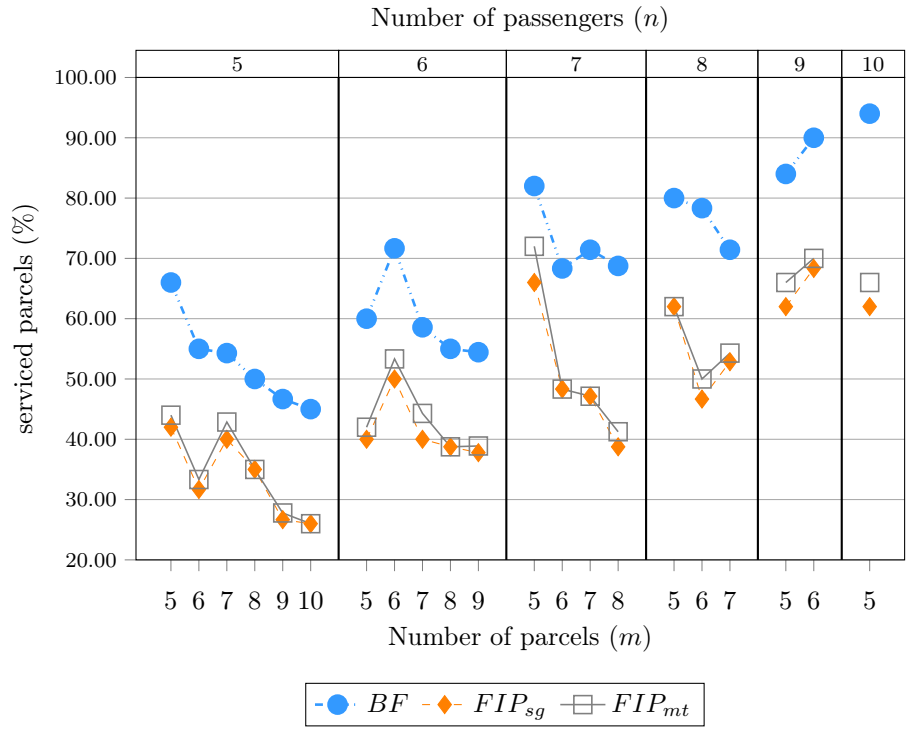

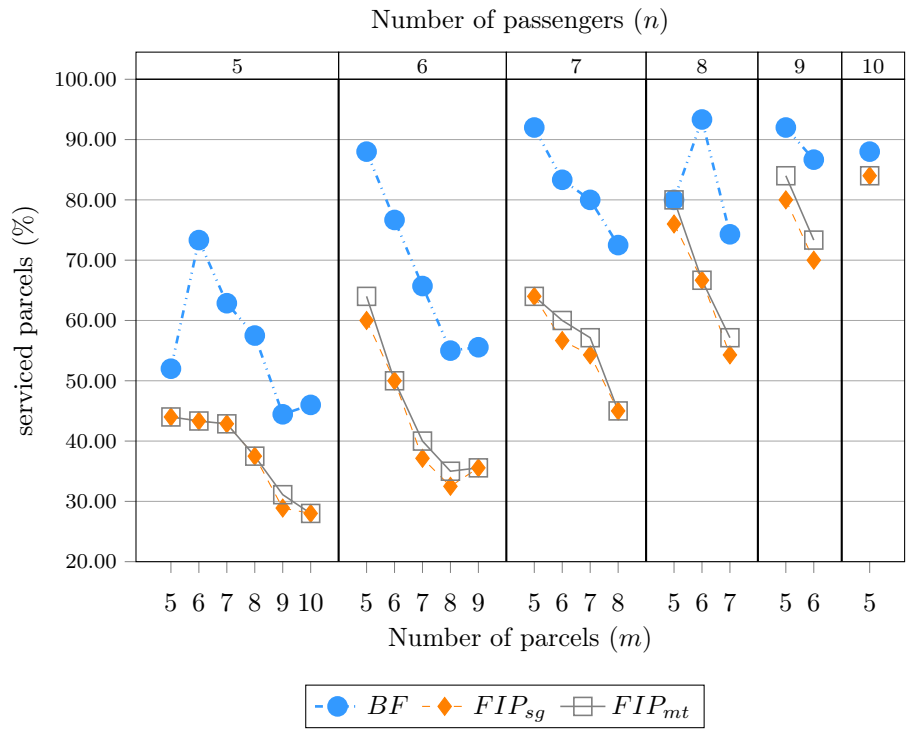
\begin{figure}[!htbp]
    \centering
    \footnotesize

    \pgfplotstableread[col sep=&, header=true]{
description     &       $BF$ &      $FIP_{sg}$ & $FIP_{mt}$

Sarp-5-5 & 52.00 & 44.00 & 44.00  
Sarp-5-6 & 73.33 & 43.33 & 43.33  
Sarp-5-7 & 62.86 & 42.86 & 42.86  
Sarp-5-8 & 57.50 & 37.50 & 37.50  
Sarp-5-9 & 44.44 & 28.89 & 31.11  
Sarp-5-10 & 46.00 & 28.00 & 28.00  
          & nan & nan & nan
Sarp-6-5 & 88.00 & 60.00 & 64.00  
Sarp-6-6 & 76.67 & 50.00 & 50.00  
Sarp-6-7 & 65.71 & 37.14 & 40.00  
Sarp-6-8 & 55.00 & 32.50 & 35.00  
Sarp-6-9 & 55.56 & 35.56 & 35.56  
          & nan & nan & nan
Sarp-7-5 & 92.00 & 64.00 & 64.00  
Sarp-7-6 & 83.33 & 56.67 & 60.00  
Sarp-7-7 & 80.00 & 54.29 & 57.14  
Sarp-7-8 & 72.50 & 45.00 & 45.00  
          & nan & nan & nan
Sarp-8-5 & 80.00 & 76.00 & 80.00  
Sarp-8-6 & 93.33 & 66.67 & 66.67  
Sarp-8-7 & 74.29 & 54.29 & 57.14  
          & nan & nan & nan
Sarp-9-5 & 92.00 & 80.00 & 84.00  
Sarp-9-6 & 86.67 & 70.00 & 73.33 
          & nan & nan & nan
Sarp-10-5 & 88.00 & 84.00 & 84.00  
}\datatableentry
    \begin{tikzpicture}
        \begin{axis}[
            height=8cm, width=11.8cm,
            ymin=20, ymax=100,
            ytick={20,30,...,100},
            grid=both,
            major grid style={line width=.1pt,draw=gray!70},
            xmajorgrids=false,
            tick align=outside,
            title style={at={(0.5,1)},anchor=south,yshift=8 },
            title={\small Number of passengers ($n$)},
            tickpos=left,
            xtick=data,
            xticklabels ={5, 6, 7, 8, 9, 10, 5, 6, 7, 8, 9, 5, 6, 7, 8, 5, 6, 7, 5, 6, 5},
            enlarge x limits = 0.05,
            yticklabel style = { 
                /pgf/number format/fixed zerofill,
                /pgf/number format/precision=2,
            },
            legend columns=5,
            legend columns=7,
            legend style={
                at={(0.48,-0.3)},
                anchor=center,
                draw=black,
                legend cell align={center},
                /tikz/every even column/.append style={font=\small, column sep=0.1cm}
            },
            ylabel={\small serviced parcels (\%)},xlabel={\small Number of parcels ($m$)},
            every x tick label/.append style={font=\small,yshift = -5 pt},
            extra description/.code={
            \draw (axis cs: -1.25,100) rectangle (axis cs: 6,104.5) node[pos=.5] {5};
            \draw (axis cs: 6,100) rectangle (axis cs: 12,104.5) node[pos=.5] {6};
            \draw (axis cs: 12,100) rectangle (axis cs: 17,104.5) node[pos=.5] {7};
            \draw (axis cs: 17,100) rectangle (axis cs: 21,104.5) node[pos=.5] {8};
            \draw (axis cs: 21,100) rectangle (axis cs: 24,104.5) node[pos=.5] {9};
            \draw (axis cs: 24,100) rectangle (axis cs: 26.25,104.5) node[pos=.5] {10};
            }
        ]
        \addplot [dashdotdotted, line width=1pt, color=color10,mark=*, unbounded coords=jump, mark options={scale=1.7,solid}] table [y=$BF$, x expr=\coordindex] {\datatableentry};
        \addplot [dashed,color=color9,mark=diamond*, unbounded coords=jump, mark options={scale=1.7,solid}] table [y=$FIP_{sg}$, x expr=\coordindex] {\datatableentry};
        \addplot [semithick,color=gray,mark=square, unbounded coords=jump, mark options={scale=1.7,solid}] table [y=$FIP_{mt}$, x expr=\coordindex] {\datatableentry};

        \addplot[thick, black, forget plot] coordinates {(6,20)(6,100)};
        \addplot[thick, black, forget plot] coordinates {(12,20)(12,100)};
        \addplot[thick, black, forget plot] coordinates {(17,20)(17,100)};        \addplot[thick, black, forget plot] coordinates {(21,20)(21,100)};
        \addplot[thick, black, forget plot] coordinates {(24,20)(24,100)};
        
    \legend{$BF$,$FIP_{sg}$,$FIP_{mt}$}
    \end{axis}
    \end{tikzpicture}
    \caption{Amount of serviced parcels (\%) for $SF_1$ dataset}
    \label{fig:servedssf}
\end{figure}

\begin{figure}[!htbp]
    \centering
    \footnotesize

    \pgfplotstableread[col sep=&, header=true]{
description     &       $BF$ &      $FIP_{sg}$ & $FIP_{mt}$
sfsarp-5-15-1 & 24.00 & 18.67 & 18.67  
sfsarp-5-20-1 & 23.00 & 13.00 & 13.00  
sfsarp-5-30-1 & 14.67 & 9.33 & 9.33  
          & nan & nan & nan
sfsarp-10-10-1 & 88.00 & 48.00 & 54.00  
sfsarp-10-15-1 & 57.33 & 34.67 & 34.67  
sfsarp-10-20-1 & 40.00 & 26.00 & 27.00  
sfsarp-10-30-1 & 30.67 & 18.00 & 18.00  
          & nan & nan & nan
sfsarp-15-10-1 & 94.00 & 60.00 & 60.00  
sfsarp-15-15-1 & 81.33 & 37.33 & 37.33  
sfsarp-15-20-1 & 65.00 & 35.00 & 37.00  
sfsarp-15-30-1 & 46.00 & 26.00 & 26.00  
          & nan & nan & nan
sfsarp-20-10-1 & 96.00 & 86.00 & 90.00  
sfsarp-20-15-1 & 93.33 & 50.67 & 50.67  
sfsarp-20-20-1 & 77.00 & 42.00 & 44.00  
sfsarp-20-30-1 & 62.00 & 32.67 & 35.33  
          & nan & nan & nan
sfsarp-30-10-1 & 98.00 & 92.00 & 92.00  
sfsarp-30-15-1 & 96.00 & 88.00 & 90.67  
sfsarp-30-20-1 & 90.00 & 68.00 & 72.00  
sfsarp-30-30-1 & 84.67 & 44.67 & 46.67 

}\datatableentry
    \begin{tikzpicture}
        \begin{axis}[
            height=8cm, width=11.8cm,
            ymin=0, ymax=100,
            ytick={0,10,...,100},
            grid=both,
            major grid style={line width=.1pt,draw=gray!70},
            xmajorgrids=false,
            tick align=outside,
            title style={at={(0.5,1)},anchor=south,yshift=8 },
            title={\small Number of passengers ($n$)},
            tickpos=left,
            xtick=data,
            xticklabels ={15, 20, 30, 10, 15, 20, 30, 10, 15, 20, 30, 10, 15, 20, 30, 10, 15, 20, 30},
            enlarge x limits = 0.05,
            yticklabel style = { 
                /pgf/number format/fixed zerofill,
                /pgf/number format/precision=2,
            },
            legend columns=5,
            legend columns=7,
            legend style={
                at={(0.48,-0.3)},
                anchor=center,
                draw=black,
                legend cell align={center},
                /tikz/every even column/.append style={font=\small, column sep=0.1cm}
            },
            ylabel={\small serviced parcels (\%)},xlabel={\small Number of parcels ($m$)},
            every x tick label/.append style={font=\small,yshift = -5 pt},
            extra description/.code={
            \draw (axis cs: -1.1,100) rectangle (axis cs: 3,105) node[pos=.5] {5};
            \draw (axis cs: 3,100) rectangle (axis cs: 8,105) node[pos=.5] {10};
            \draw (axis cs: 8,100) rectangle (axis cs: 13,105) node[pos=.5] {15};
            \draw (axis cs: 13,100) rectangle (axis cs: 18,105) node[pos=.5] {20};
            \draw (axis cs: 18,100) rectangle (axis cs: 23.1,105) node[pos=.5] {30};
            }
        ]
        \addplot [dashdotdotted, line width=1pt, color=color10,mark=*, unbounded coords=jump, mark options={scale=1.7,solid}] table [y=$BF$, x expr=\coordindex] {\datatableentry};
        \addplot [dashed,color=color9,mark=diamond*, unbounded coords=jump, mark options={scale=1.7,solid}] table [y=$FIP_{sg}$, x expr=\coordindex] {\datatableentry};
        \addplot [semithick,color=gray,mark=square, unbounded coords=jump, mark options={scale=1.7,solid}] table [y=$FIP_{mt}$, x expr=\coordindex] {\datatableentry};

        \addplot[thick, black, forget plot] coordinates {(3,0)(3,100)};
        \addplot[thick, black, forget plot] coordinates {(8,0)(8,100)};
        \addplot[thick, black, forget plot] coordinates {(13,0)(13,100)};        \addplot[thick, black, forget plot] coordinates {(18,0)(18,100)};
        
    \legend{$BF$,$FIP_{sg}$,$FIP_{mt}$}
    \end{axis}
    \end{tikzpicture}
    \caption{Amount of serviced parcels (\%) for $SF_2$ dataset}
    \label{fig:servedmsf}
\end{figure}
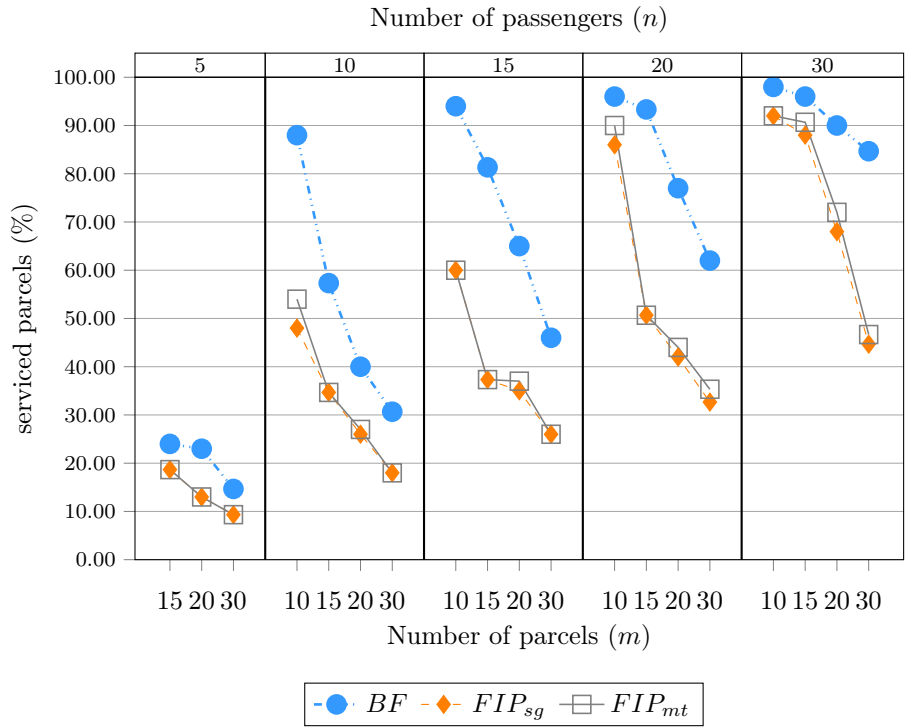

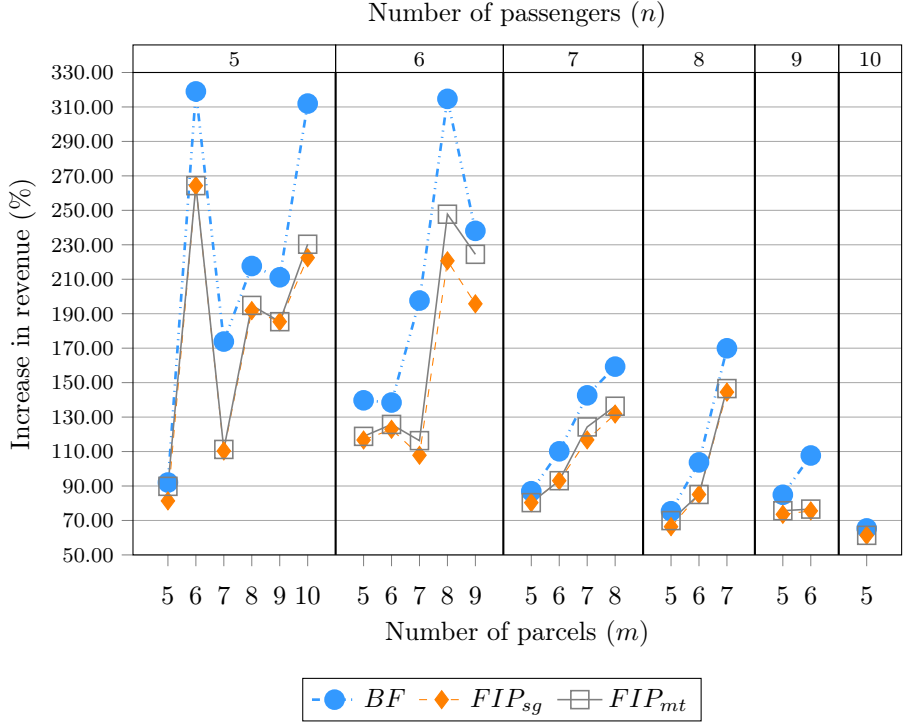
\begin{figure}[!htbp]
    \centering
    \footnotesize

    \pgfplotstableread[col sep=&, header=true]{
description     &       $BF$ &      $FIP_{sg}$ & $FIP_{mt}$

Sarp-5-5 & 92.02 & 81.35 & 89.80  
Sarp-5-6 & 319.02 & 264.33 & 264.33  
Sarp-5-7 & 173.84 & 110.32 & 111.23  
Sarp-5-8 & 217.68 & 191.81 & 194.82  
Sarp-5-9 & 211.13 & 185.37 & 185.37  
Sarp-5-10 & 311.97 & 222.48 & 230.32
          & nan & nan & nan
Sarp-6-5 & 139.73 & 116.69 & 118.81  
Sarp-6-6 & 138.45 & 122.91 & 125.74  
Sarp-6-7 & 197.63 & 107.85 & 116.33  
Sarp-6-8 & 314.70 & 220.70 & 247.74  
Sarp-6-9 & 238.09 & 195.74 & 224.47  
          & nan & nan & nan
Sarp-7-5 & 86.97 & 80.35 & 80.35  
Sarp-7-6 & 110.13 & 93.08 & 93.08  
Sarp-7-7 & 142.62 & 116.76 & 124.29  
Sarp-7-8 & 159.29 & 131.77 & 136.30  
          & nan & nan & nan
Sarp-8-5 & 75.35 & 66.41 & 69.92  
Sarp-8-6 & 103.71 & 85.08 & 85.08  
Sarp-8-7 & 169.93 & 144.49 & 146.55  
          & nan & nan & nan
Sarp-9-5 & 84.90 & 73.55 & 75.65  
Sarp-9-6 & 107.71 & 75.56 & 76.60  
          & nan & nan & nan
Sarp-10-5 & 65.39 & 61.37 & 61.37  
}\datatableentry
    \begin{tikzpicture}
        \begin{axis}[
            height=8cm, width=11.8cm,
            ymin=50, ymax=330,
            ytick={50,70,...,330},
            grid=both,
            major grid style={line width=.1pt,draw=gray!70},
            xmajorgrids=false,
            tick align=outside,
            title style={at={(0.5,1)},anchor=south,yshift=8 },
            title={\small Number of passengers ($n$)},
            tickpos=left,
            xtick=data,
            xticklabels ={5, 6, 7, 8, 9, 10, 5, 6, 7, 8, 9, 5, 6, 7, 8, 5, 6, 7, 5, 6, 5},
            enlarge x limits = 0.05,
            yticklabel style = { 
                /pgf/number format/fixed zerofill,
                /pgf/number format/precision=2,
            },
            legend columns=5,
            legend columns=7,
            legend style={
                at={(0.48,-0.3)},
                anchor=center,
                draw=black,
                legend cell align={center},
                /tikz/every even column/.append style={font=\small, column sep=0.1cm}
            },
            ylabel={\small Increase in revenue (\%)},xlabel={\small Number of parcels ($m$)},
            every x tick label/.append style={font=\small,yshift = -5 pt},
            extra description/.code={
            \draw (axis cs: -1.25,330) rectangle (axis cs: 6,346) node[pos=.5] {5};
            \draw (axis cs: 6,330) rectangle (axis cs: 12,346) node[pos=.5] {6};
            \draw (axis cs: 12,330) rectangle (axis cs: 17,346) node[pos=.5] {7};
            \draw (axis cs: 17,330) rectangle (axis cs: 21,346) node[pos=.5] {8};
            \draw (axis cs: 21,330) rectangle (axis cs: 24,346) node[pos=.5] {9};
            \draw (axis cs: 24,330) rectangle (axis cs: 26.25,346) node[pos=.5] {10};
            }
        ]
        \addplot [dashdotdotted, line width=1pt, color=color10,mark=*, unbounded coords=jump, mark options={scale=1.7,solid}] table [y=$BF$, x expr=\coordindex] {\datatableentry};
        \addplot [dashed,color=color9,mark=diamond*, unbounded coords=jump, mark options={scale=1.7,solid}] table [y=$FIP_{sg}$, x expr=\coordindex] {\datatableentry};
        \addplot [semithick,color=gray,mark=square, unbounded coords=jump, mark options={scale=1.7,solid}] table [y=$FIP_{mt}$, x expr=\coordindex] {\datatableentry};

        \addplot[thick, black, forget plot] coordinates {(6,50)(6,330)};
        \addplot[thick, black, forget plot] coordinates {(12,50)(12,330)};
        \addplot[thick, black, forget plot] coordinates {(17,50)(17,330)};        \addplot[thick, black, forget plot] coordinates {(21,50)(21,330)};
        \addplot[thick, black, forget plot] coordinates {(24,50)(24,330)};
        
    \legend{$BF$,$FIP_{sg}$,$FIP_{mt}$}
    \end{axis}
    \end{tikzpicture}
    \caption{Increase in revenue (\%) for $MD_1$ dataset}
    \label{fig:revsmd}
\end{figure}

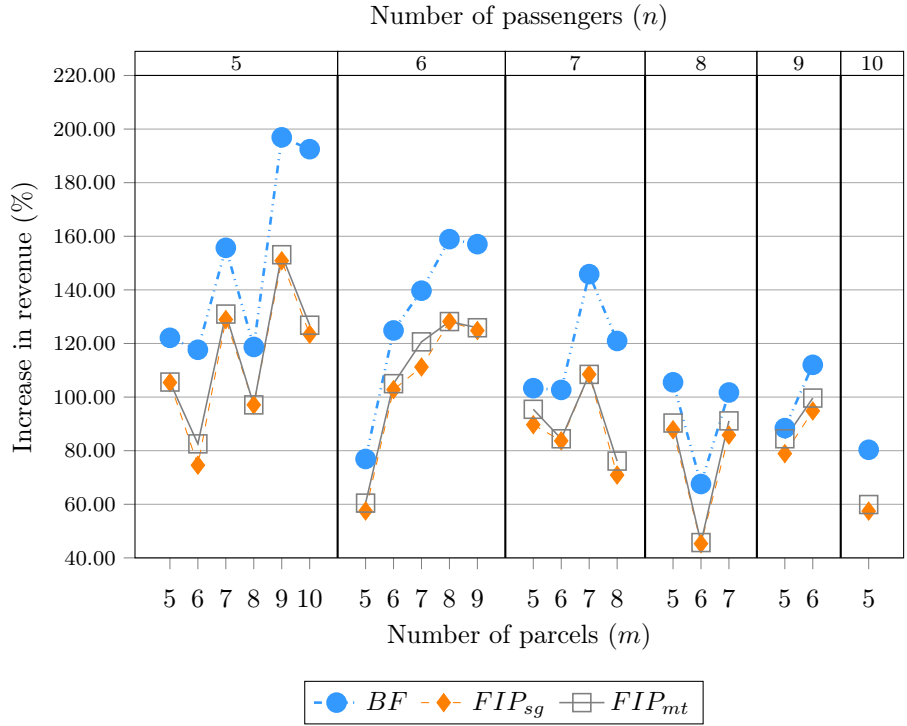
\begin{figure}[!htbp]
    \centering
    \footnotesize

    \pgfplotstableread[col sep=&, header=true]{
description     &       $BF$ &      $FIP_{sg}$ & $FIP_{mt}$

Sarp-5-5 & 122.11 & 105.43 & 105.61  
Sarp-5-6 & 117.65 & 74.58 & 82.55  
Sarp-5-7 & 155.69 & 128.93 & 130.98  
Sarp-5-8 & 118.71 & 97.09 & 97.09  
Sarp-5-9 & 196.91 & 150.88 & 153.12  
Sarp-5-10 & 192.50 & 123.42 & 126.82  
          & nan & nan & nan
Sarp-6-5 & 76.91 & 57.48 & 60.44  
Sarp-6-6 & 124.86 & 102.77 & 105.01  
Sarp-6-7 & 139.70 & 111.21 & 120.61  
Sarp-6-8 & 158.93 & 128.18 & 128.18  
Sarp-6-9 & 157.06 & 124.83 & 125.87  
          & nan & nan & nan
Sarp-7-5 & 103.32 & 89.70 & 95.44  
Sarp-7-6 & 102.68 & 83.67 & 84.44  
Sarp-7-7 & 145.87 & 108.47 & 108.47  
Sarp-7-8 & 120.94 & 70.90 & 76.12  
          & nan & nan & nan
Sarp-8-5 & 105.52 & 87.82 & 90.32  
Sarp-8-6 & 67.57 & 45.31 & 45.72  
Sarp-8-7 & 101.71 & 85.81 & 91.12  
          & nan & nan & nan
Sarp-9-5 & 88.42 & 78.91 & 84.45  
Sarp-9-6 & 112.03 & 94.82 & 99.63  
          & nan & nan & nan
Sarp-10-5 & 80.34 & 57.45 & 59.91   
}\datatableentry
    \begin{tikzpicture}
        \begin{axis}[
            height=8cm, width=11.8cm,
            ymin=40, ymax=220,
            ytick={40,60,...,220},
            grid=both,
            major grid style={line width=.1pt,draw=gray!70},
            xmajorgrids=false,
            tick align=outside,
            title style={at={(0.5,1)},anchor=south,yshift=8 },
            title={\small Number of passengers ($n$)},
            tickpos=left,
            xtick=data,
            xticklabels ={5, 6, 7, 8, 9, 10, 5, 6, 7, 8, 9, 5, 6, 7, 8, 5, 6, 7, 5, 6, 5},
            enlarge x limits = 0.05,
            yticklabel style = { 
                /pgf/number format/fixed zerofill,
                /pgf/number format/precision=2,
            },
            legend columns=5,
            legend columns=7,
            legend style={
                at={(0.48,-0.3)},
                anchor=center,
                draw=black,
                legend cell align={center},
                /tikz/every even column/.append style={font=\small, column sep=0.1cm}
            },
            ylabel={\small Increase in revenue (\%)},xlabel={\small Number of parcels ($m$)},
            every x tick label/.append style={font=\small,yshift = -5 pt},
            extra description/.code={
            \draw (axis cs: -1.25,220) rectangle (axis cs: 6,229) node[pos=.5] {5};
            \draw (axis cs: 6,220) rectangle (axis cs: 12,229) node[pos=.5] {6};
            \draw (axis cs: 12,220) rectangle (axis cs: 17,229) node[pos=.5] {7};
            \draw (axis cs: 17,220) rectangle (axis cs: 21,229) node[pos=.5] {8};
            \draw (axis cs: 21,220) rectangle (axis cs: 24,229) node[pos=.5] {9};
            \draw (axis cs: 24,220) rectangle (axis cs: 26.25,229) node[pos=.5] {10};
            }
        ]
        \addplot [dashdotdotted, line width=1pt, color=color10,mark=*, unbounded coords=jump, mark options={scale=1.7,solid}] table [y=$BF$, x expr=\coordindex] {\datatableentry};
        \addplot [dashed,color=color9,mark=diamond*, unbounded coords=jump, mark options={scale=1.7,solid}] table [y=$FIP_{sg}$, x expr=\coordindex] {\datatableentry};
        \addplot [semithick,color=gray,mark=square, unbounded coords=jump, mark options={scale=1.7,solid}] table [y=$FIP_{mt}$, x expr=\coordindex] {\datatableentry};

        \addplot[thick, black, forget plot] coordinates {(6,40)(6,220)};
        \addplot[thick, black, forget plot] coordinates {(12,40)(12,220)};
        \addplot[thick, black, forget plot] coordinates {(17,40)(17,220)};        \addplot[thick, black, forget plot] coordinates {(21,40)(21,220)};
        \addplot[thick, black, forget plot] coordinates {(24,40)(24,220)};
        
    \legend{$BF$,$FIP_{sg}$,$FIP_{mt}$}
    \end{axis}
    \end{tikzpicture}
    \caption{Increase in revenue (\%) for $GH_1$ dataset}
    \label{fig:revsgh}
\end{figure}

\begin{figure}[!htbp]
    \centering
    \footnotesize

    \pgfplotstableread[col sep=&, header=true]{
description     &       $BF$ &      $FIP_{sg}$ & $FIP_{mt}$
Sarp-5-5 & 74.08 & 63.40 & 68.51  
Sarp-5-6 & 83.10 & 58.03 & 59.96  
Sarp-5-7 & 102.30 & 77.55 & 81.95  
Sarp-5-8 & 86.41 & 62.28 & 67.56  
Sarp-5-9 & 68.37 & 53.03 & 55.56  
Sarp-5-10 & 73.31 & 51.27 & 52.25  
          & nan & nan & nan
Sarp-6-5 & 91.46 & 75.83 & 77.28  
Sarp-6-6 & 79.59 & 61.36 & 64.78  
Sarp-6-7 & 62.05 & 42.46 & 46.74  
Sarp-6-8 & 58.63 & 41.01 & 43.42  
Sarp-6-9 & 105.77 & 71.66 & 75.97  
          & nan & nan & nan
Sarp-7-5 & 66.53 & 47.59 & 50.46  
Sarp-7-6 & 72.45 & 55.81 & 61.88  
Sarp-7-7 & 80.01 & 61.49 & 66.35  
Sarp-7-8 & 69.88 & 49.73 & 51.90 
          & nan & nan & nan
Sarp-8-5 & 47.98 & 44.15 & 45.62  
Sarp-8-6 & 94.74 & 74.67 & 75.90  
Sarp-8-7 & 60.97 & 47.28 & 50.01  
          & nan & nan & nan
Sarp-9-5 & 42.80 & 40.08 & 41.07  
Sarp-9-6 & 42.12 & 35.40 & 36.86 
          & nan & nan & nan
Sarp-10-5 & 38.85 & 34.96 & 37.40 
}\datatableentry
    \begin{tikzpicture}
        \begin{axis}[
            height=8cm, width=11.8cm,
            ymin=30, ymax=110,
            ytick={30,40,...,110},
            grid=both,
            major grid style={line width=.1pt,draw=gray!70},
            xmajorgrids=false,
            tick align=outside,
            title style={at={(0.5,1)},anchor=south,yshift=8 },
            title={\small Number of passengers ($n$)},
            tickpos=left,
            xtick=data,
            xticklabels ={5, 6, 7, 8, 9, 10, 5, 6, 7, 8, 9, 5, 6, 7, 8, 5, 6, 7, 5, 6, 5},
            enlarge x limits = 0.05,
            yticklabel style = { 
                /pgf/number format/fixed zerofill,
                /pgf/number format/precision=2,
            },
            legend columns=5,
            legend columns=7,
            legend style={
                at={(0.48,-0.3)},
                anchor=center,
                draw=black,
                legend cell align={center},
                /tikz/every even column/.append style={font=\small, column sep=0.1cm}
            },
            ylabel={\small Increase in revenue (\%)},xlabel={\small Number of parcels ($m$)},
            every x tick label/.append style={font=\small,yshift = -5 pt},
            extra description/.code={
            \draw (axis cs: -1.25,110) rectangle (axis cs: 6,114) node[pos=.5] {5};
            \draw (axis cs: 6,110) rectangle (axis cs: 12,114) node[pos=.5] {6};
            \draw (axis cs: 12,110) rectangle (axis cs: 17,114) node[pos=.5] {7};
            \draw (axis cs: 17,110) rectangle (axis cs: 21,114) node[pos=.5] {8};
            \draw (axis cs: 21,110) rectangle (axis cs: 24,114) node[pos=.5] {9};
            \draw (axis cs: 24,110) rectangle (axis cs: 26.25,114) node[pos=.5] {10};
            }
        ]
        \addplot [dashdotdotted, line width=1pt, color=color10,mark=*, unbounded coords=jump, mark options={scale=1.7,solid}] table [y=$BF$, x expr=\coordindex] {\datatableentry};
        \addplot [dashed,color=color9,mark=diamond*, unbounded coords=jump, mark options={scale=1.7,solid}] table [y=$FIP_{sg}$, x expr=\coordindex] {\datatableentry};
        \addplot [semithick,color=gray,mark=square, unbounded coords=jump, mark options={scale=1.7,solid}] table [y=$FIP_{mt}$, x expr=\coordindex] {\datatableentry};

        \addplot[thick, black, forget plot] coordinates {(6,30)(6,110)};
        \addplot[thick, black, forget plot] coordinates {(12,30)(12,110)};
        \addplot[thick, black, forget plot] coordinates {(17,30)(17,110)};        \addplot[thick, black, forget plot] coordinates {(21,30)(21,110)};
        \addplot[thick, black, forget plot] coordinates {(24,30)(24,110)};
        
    \legend{$BF$,$FIP_{sg}$,$FIP_{mt}$}
    \end{axis}
    \end{tikzpicture}
    \caption{Increase in revenue (\%) for $SF_1$ dataset}
    \label{fig:revssf}
\end{figure}
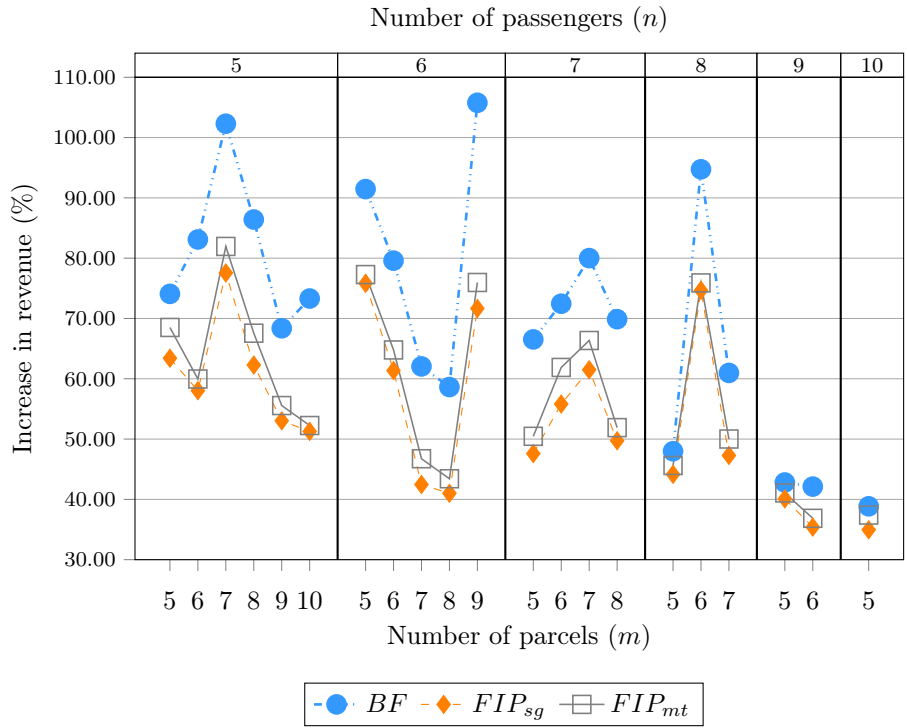

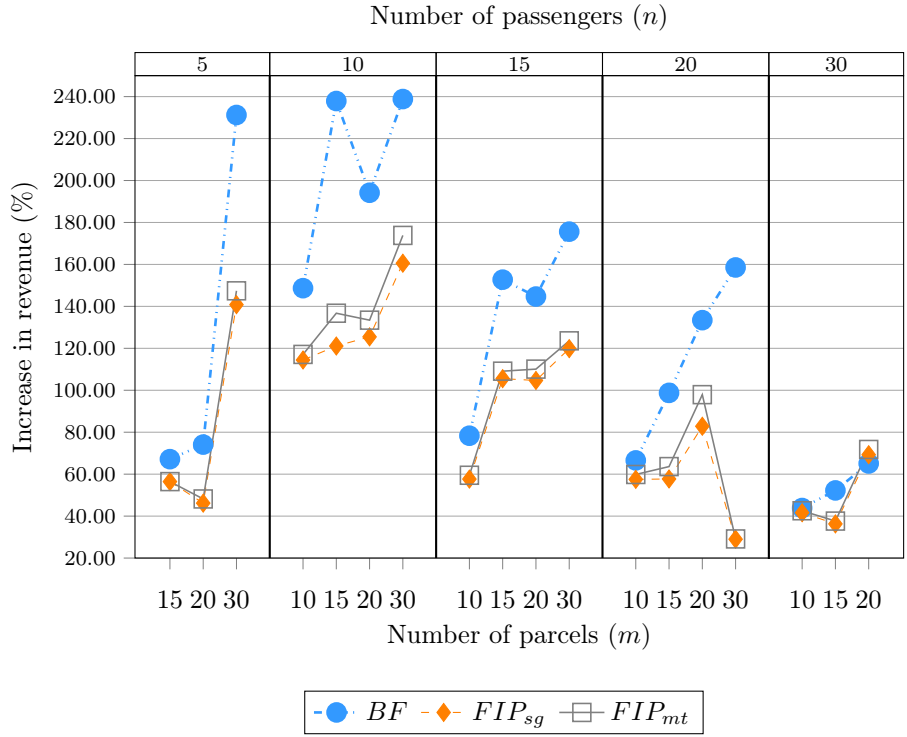
\begin{figure}[!htbp]
    \centering
    \footnotesize

    \pgfplotstableread[col sep=&, header=true]{
description     &       $BF$ &      $FIP_{sg}$ & $FIP_{mt}$
sarp-5-15-A-1 & 67.15 & 56.45 & 56.45  
sarp-5-20-A-1 & 74.10 & 46.18 & 48.18
sarp-5-30-A-1 & 231.24 & 140.80 & 147.41 
          & nan & nan & nan
sarp-10-10-A-1 & 148.67 & 114.42 & 117.09  
sarp-10-15-A-1 & 237.90 & 121.09 & 136.73  
sarp-10-20-A-1 & 194.17 & 125.46 & 133.42  
sarp-10-30-A-1 & 238.84 & 160.57 & 173.83
          & nan & nan & nan
sarp-15-10-A-1 & 78.33 & 57.65 & 59.48  
sarp-15-15-A-1 & 152.73 & 105.54 & 109.10  
sarp-15-20-A-1 & 144.68 & 104.67 & 110.08  
sarp-15-30-A-1 & 175.62 & 119.85 & 123.57  
          & nan & nan & nan
sarp-20-10-A-1 & 66.56 & 57.51 & 59.85  
sarp-20-15-A-1 & 98.76 & 57.71 & 63.61  
sarp-20-20-A-1 & 133.42 & 82.84 & 97.91  
sarp-20-30-A-1 & 158.55 & 28.99 & 29.12  
          & nan & nan & nan
sarp-30-10-A-1 & 43.85 & 41.63 & 42.55  
sarp-30-15-A-1 & 52.25 & 36.28 & 37.63  
sarp-30-20-A-1 & 65.09 & 69.28 & 71.78

}\datatableentry
    \begin{tikzpicture}
        \begin{axis}[
            height=8cm, width=11.8cm,
            ymin=20, ymax=250,
            ytick={20,40,...,250},
            grid=both,
            major grid style={line width=.1pt,draw=gray!70},
            xmajorgrids=false,
            tick align=outside,
            title style={at={(0.5,1)},anchor=south,yshift=8 },
            title={\small Number of passengers ($n$)},
            tickpos=left,
            xtick=data,
            xticklabels ={15, 20, 30, 10, 15, 20, 30, 10, 15, 20, 30, 10, 15, 20, 30, 10, 15, 20},
            enlarge x limits = 0.05,
            yticklabel style = { 
                /pgf/number format/fixed zerofill,
                /pgf/number format/precision=2,
            },
            legend columns=5,
            legend columns=7,
            legend style={
                at={(0.48,-0.32)},
                anchor=center,
                draw=black,
                legend cell align={center},
                /tikz/every even column/.append style={font=\small, column sep=0.1cm}
            },
            ylabel={\small Increase in revenue (\%)},xlabel={\small Number of parcels ($m$)},
            every x tick label/.append style={font=\small,yshift = -5 pt},
            extra description/.code={
            \draw (axis cs: -1.05,250) rectangle (axis cs: 3,261) node[pos=.5] {5};
            \draw (axis cs: 3,250) rectangle (axis cs: 8,261) node[pos=.5] {10};
            \draw (axis cs: 8,250) rectangle (axis cs: 13,261) node[pos=.5] {15};
            \draw (axis cs: 13,250) rectangle (axis cs: 18,261) node[pos=.5] {20};
            \draw (axis cs: 18,250) rectangle (axis cs: 22.05,261) node[pos=.5] {30};
             }
        ]
        \addplot [dashdotdotted, line width=1pt, color=color10,mark=*, unbounded coords=jump, mark options={scale=1.7,solid}] table [y=$BF$, x expr=\coordindex] {\datatableentry};
        \addplot [dashed,color=color9,mark=diamond*, unbounded coords=jump, mark options={scale=1.7,solid}] table [y=$FIP_{sg}$, x expr=\coordindex] {\datatableentry};
        \addplot [semithick,color=gray,mark=square, unbounded coords=jump, mark options={scale=1.7,solid}] table [y=$FIP_{mt}$, x expr=\coordindex] {\datatableentry};

        \addplot[thick, black, forget plot] coordinates {(3,20)(3,250)};
        \addplot[thick, black, forget plot] coordinates {(8,20)(8,250)};
        \addplot[thick, black, forget plot] coordinates {(13,20)(13,250)};     
        \addplot[thick, black, forget plot] coordinates {(18,0)(18,250)};
        
    \legend{$BF$,$FIP_{sg}$,$FIP_{mt}$}
    \end{axis}
    \end{tikzpicture}
    \caption{Increase in revenue (\%) for $MD_2$ dataset}
    \label{fig:revmmd}
\end{figure}

\begin{figure}[!htbp]
    \centering
    \footnotesize

    \pgfplotstableread[col sep=&, header=true]{
description     &       $BF$ &      $FIP_{sg}$ & $FIP_{mt}$
sfsarp-5-15-1 & 78.90 & 65.26 & 66.85  
sfsarp-5-20-1 & 108.80 & 74.18 & 75.75  
sfsarp-5-30-1 & 122.07 & 87.79 & 89.81 
          & nan & nan & nan
sfsarp-10-10-1 & 104.10 & 69.04 & 74.00  
sfsarp-10-15-1 & 109.93 & 77.35 & 81.95  
sfsarp-10-20-1 & 81.62 & 57.33 & 60.57  
sfsarp-10-30-1 & 107.67 & 77.60 & 84.05  
          & nan & nan & nan
sfsarp-15-10-1 & 57.50 & 38.89 & 40.08  
sfsarp-15-15-1 & 67.06 & 35.40 & 37.24  
sfsarp-15-20-1 & 81.91 & 56.96 & 60.80  
sfsarp-15-30-1 & 92.17 & 61.29 & 63.82  
          & nan & nan & nan
sfsarp-20-10-1 & 54.59 & 48.37 & 50.64  
sfsarp-20-15-1 & 66.48 & 38.23 & 41.25  
sfsarp-20-20-1 & 67.51 & 44.75 & 49.83  
sfsarp-20-30-1 & 100.57 & 62.15 & 67.88  
          & nan & nan & nan
sfsarp-30-10-1 & 40.65 & 37.11 & 38.74  
sfsarp-30-15-1 & 51.25 & 46.08 & 48.81  
sfsarp-30-20-1 & 57.72 & 48.40 & 51.85  
sfsarp-30-30-1 & 79.14 & 51.83 & 55.99  

}\datatableentry
    \begin{tikzpicture}
        \begin{axis}[
            height=8cm, width=11.8cm,
            ymin=20, ymax=130,
            ytick={0,20,...,130},
            grid=both,
            major grid style={line width=.1pt,draw=gray!70},
            xmajorgrids=false,
            tick align=outside,
            title style={at={(0.5,1)},anchor=south,yshift=8 },
            title={\small Number of passengers ($n$)},
            tickpos=left,
            xtick=data,
            xticklabels ={15, 20, 30, 10, 15, 20, 30, 10, 15, 20, 30, 10, 15, 20, 30, 10, 15, 20, 30},
            enlarge x limits = 0.05,
            yticklabel style = { 
                /pgf/number format/fixed zerofill,
                /pgf/number format/precision=2,
            },
            legend columns=5,
            legend columns=7,
            legend style={
                at={(0.48,-0.32)},
                anchor=center,
                draw=black,
                legend cell align={center},
                /tikz/every even column/.append style={font=\small, column sep=0.1cm}
            },
            ylabel={\small Increase in revenue (\%)},xlabel={\small Number of parcels ($m$)},
            every x tick label/.append style={font=\small,yshift = -5 pt},
            extra description/.code={
            \draw (axis cs: -1.1,130) rectangle (axis cs: 3,135.5) node[pos=.5] {5};
            \draw (axis cs: 3,130) rectangle (axis cs: 8,135.5) node[pos=.5] {10};
            \draw (axis cs: 8,130) rectangle (axis cs: 13,135.5) node[pos=.5] {15};
            \draw (axis cs: 13,130) rectangle (axis cs: 18,135.5) node[pos=.5] {20};
            \draw (axis cs: 18,130) rectangle (axis cs: 23.1,135.5) node[pos=.5] {30};
            }
        ]
        \addplot [dashdotdotted, line width=1pt, color=color10,mark=*, unbounded coords=jump, mark options={scale=1.7,solid}] table [y=$BF$, x expr=\coordindex] {\datatableentry};
        \addplot [dashed,color=color9,mark=diamond*, unbounded coords=jump, mark options={scale=1.7,solid}] table [y=$FIP_{sg}$, x expr=\coordindex] {\datatableentry};
        \addplot [semithick,color=gray,mark=square, unbounded coords=jump, mark options={scale=1.7,solid}] table [y=$FIP_{mt}$, x expr=\coordindex] {\datatableentry};

        \addplot[thick, black, forget plot] coordinates {(3,20)(3,130)};
        \addplot[thick, black, forget plot] coordinates {(8,20)(8,130)};
        \addplot[thick, black, forget plot] coordinates {(13,20)(13,130)};        \addplot[thick, black, forget plot] coordinates {(18,20)(18,130)};
        
    \legend{$BF$,$FIP_{sg}$,$FIP_{mt}$}
    \end{axis}
    \end{tikzpicture}
    \caption{Increase in revenue (\%) for $SF_2$ dataset}
    \label{fig:revmsf}
\end{figure}
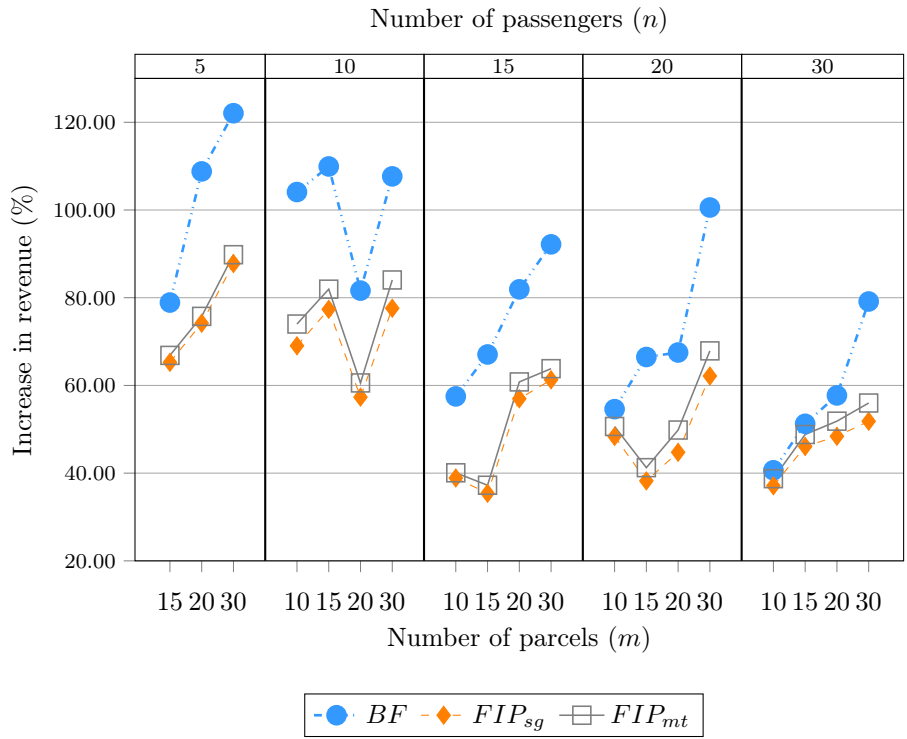

Two parameters were used to compare cost and service aspects between formulations. The first one concerns the portion of available parcel requests that are serviced and its results are depicted in Figures \ref{fig:servedsmd}--\ref{fig:servedmsf}. The second parameter measures the increase in profit (objective function value) obtained by adding parcel service to a passenger-exclusive system. Specifically, we calculate the percentage value offset between the tested formulations and a passenger-only solution. The results for this parameter are displayed in Figures \ref{fig:revsmd}--\ref{fig:revmsf}.

It is worth noting that the results for $ n = 5$ were scaled down to a 1/5 in Figure \ref{fig:revmmd} to better fit the order of magnitude of the other instances in the same set.

\begin{figure}[!htbp]
    \centering
    \footnotesize

    \pgfplotstableread[col sep=&, header=true]{
description     &       $BF$ &      $FIP_{sg}$ & $FIP_{mt}$
sarp-5-15-A-1 & 27.78 & 20.00 & 20.00  
sarp-5-20-A-1 & 24.17 & 13.33 & 13.33  
sarp-5-30-A-1 & 15.56 & 9.44 & 9.44  
          & nan & nan & nan
sarp-10-10-A-1 & 68.33 & 45.00 & 50.00  
sarp-10-15-A-1 & 57.78 & 23.33 & 25.56  
sarp-10-20-A-1 & 40.83 & 20.83 & 20.83  
sarp-10-30-A-1 & 31.11 & 18.89 & 19.44  
          & nan & nan & nan
sarp-15-10-A-1 & 75.00 & 48.33 & 51.67  
sarp-15-15-A-1 & 67.78 & 45.56 & 46.67  
sarp-15-20-A-1 & 61.67 & 38.33 & 40.00  
sarp-15-30-A-1 & 38.33 & 23.89 & 25.00  
          & nan & nan & nan
sarp-20-10-A-1 & 85.00 & 66.67 & 70.00  
sarp-20-15-A-1 & 81.11 & 45.56 & 47.78  
sarp-20-20-A-1 & 75.83 & 32.50 & 34.17  
sarp-20-30-A-1 & 56.11 & 13.89 & 13.89  
          & nan & nan & nan
sarp-30-10-A-1 & 93.33 & 73.33 & 76.67  
sarp-30-15-A-1 & 83.33 & 54.44 & 56.67  
sarp-30-20-A-1 & 77.00 & 59.00 & 61.00

}\datatableentry
    \begin{tikzpicture}
        \begin{axis}[
            height=8cm, width=11.8cm,
            ymin=0, ymax=100,
            ytick={0,10,...,100},
            grid=both,
            major grid style={line width=.1pt,draw=gray!70},
            xmajorgrids=false,
            tick align=outside,
            title style={at={(0.5,1)},anchor=south,yshift=8 },
            title={\small Number of passengers ($n$)},
            tickpos=left,
            xtick=data,
            xticklabels ={15, 20, 30, 10, 15, 20, 30, 10, 15, 20, 30, 10, 15, 20, 30, 10, 15, 20},
            enlarge x limits = 0.05,
            yticklabel style = { 
                /pgf/number format/fixed zerofill,
                /pgf/number format/precision=2,
            },
            legend columns=5,
            legend columns=7,
            legend style={
                at={(0.48,-0.3)},
                anchor=center,
                draw=black,
                legend cell align={center},
                /tikz/every even column/.append style={font=\small, column sep=0.1cm}
            },
            ylabel={\small serviced parcels (\%)},xlabel={\small Number of parcels ($m$)},
            every x tick label/.append style={font=\small,yshift = -5 pt},
            extra description/.code={
            \draw (axis cs: -1.05,100) rectangle (axis cs: 3,105) node[pos=.5] {5};
            \draw (axis cs: 3,100) rectangle (axis cs: 8,105) node[pos=.5] {10};
            \draw (axis cs: 8,100) rectangle (axis cs: 13,105) node[pos=.5] {15};
            \draw (axis cs: 13,100) rectangle (axis cs: 18,105) node[pos=.5] {20};
            \draw (axis cs: 18,100) rectangle (axis cs: 22.05,105) node[pos=.5] {30};
            }
        ]
        \addplot [dashdotdotted, line width=1pt, color=color10,mark=*, unbounded coords=jump, mark options={scale=1.7,solid}] table [y=$BF$, x expr=\coordindex] {\datatableentry};
        \addplot [dashed,color=color9,mark=diamond*, unbounded coords=jump, mark options={scale=1.7,solid}] table [y=$FIP_{sg}$, x expr=\coordindex] {\datatableentry};
        \addplot [semithick,color=gray,mark=square, unbounded coords=jump, mark options={scale=1.7,solid}] table [y=$FIP_{mt}$, x expr=\coordindex] {\datatableentry};

        \addplot[thick, black, forget plot] coordinates {(3,0)(3,100)};
        \addplot[thick, black, forget plot] coordinates {(8,0)(8,100)};
        \addplot[thick, black, forget plot] coordinates {(13,0)(13,100)};        \addplot[thick, black, forget plot] coordinates {(18,0)(18,100)};
        
    \legend{$BF$,$FIP_{sg}$,$FIP_{mt}$}
    \end{axis}
    \end{tikzpicture}
    \caption{Amount of serviced parcels (\%) for $MD_2$ dataset}
    \label{fig:servedmmd}
\end{figure}
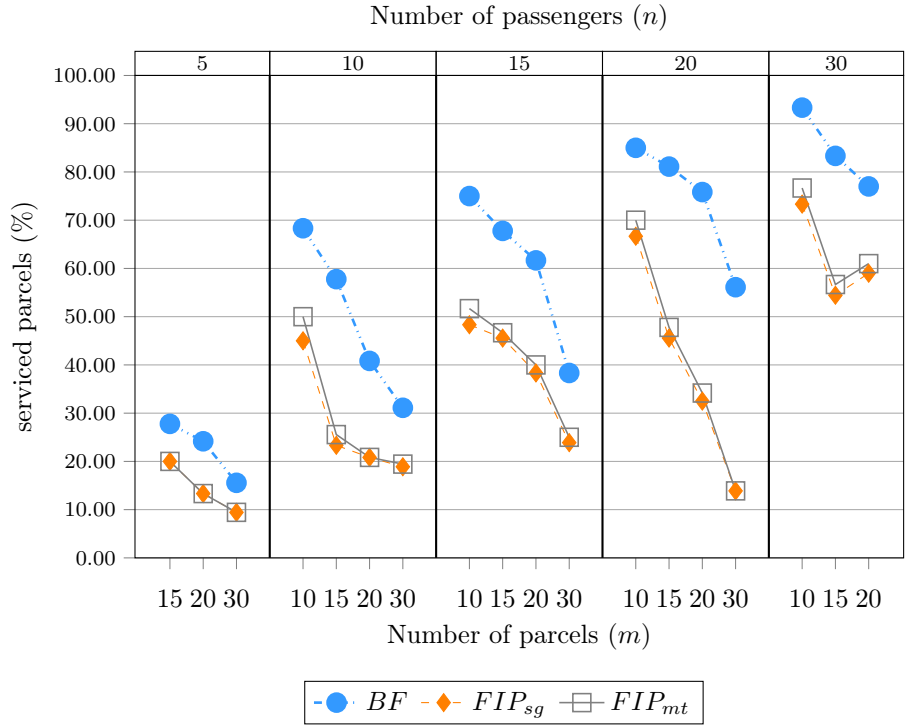

At first glance, it is possible to observe that the $BF$ consistently performs better than $FIP_{sg}$ for all tested instances. When comparing our proposed method to $FIP_{mt}$, there is also clear superiority in the results obtained by the $BF$ in most cases. Taking a general look at the performances of the methods, we can calculate that the $BF$ serviced 67\% of parcels over all instances of class $1$, compared to 47.45\% and 49.27\% for $FIP_{sg}$ and $FIP_{mt}$, respectively. In two cases for this class, namely $MD_1$-$8$-$5$ and $SF_1$-$8$-$5$, $FIP_{mt}$ and $BF$ yielded equivalent results. 
The same trend is observed in instances of class $2$, in which the $BF$ serviced 62.63\% of parcels over all instances, in contrast to 39.96\% and 41.48\% for $FIP_{sg}$ and $FIP_{mt}$, respectively. Nevertheless, it is worth noting that, in this case, neither version of FIP could obtain equivalent results to the ones provided by the BF.

Both versions of FIP provide similar results overall with better outcomes from $FIP_{mt}$ in 57.14\% and 59.80\% of instances for classes $1$ and $2$, respectively, when compared to $FIP_{sg}$. However, this difference in favour of $FIP_{mt}$ only amounts to an average increase in parcel services of 1.82\% and 1.52\% for instances of classes $1$ and $2$, respectively.

It is also worth noting that, due to capacity restrictions in the BF, the maximum amount of parcels that can be serviced in this setting is equal to the number of passenger requests available. This explains a general downward trend in the graphs representing the $BF$ results, especially in instances of class $2$.

Carrying our analysis further, we can focus on instances in which the number of passengers outweighs the number of parcels ($m\leq n$), in which the possibilities for parcel service are the most favourable. In all of them, at least 50\% of the available parcels are serviced using the BF. In 79.17\% of instances with this characteristic, the $BF$ was able to service at least 70\% of all parcel requests.

Our proposed method performs better when calculating the revenue increase from a passenger-only solution across all instances. The $BF$ improves this value in 119.86\% for instances of class $1$, compared to 93.73\% and 97.48\% obtained respectively by $FIP_{sg}$ and $FIP_{mt}$. In instances of class $2$, there is a similar trend between the solving strategies considered, in which the $BF$ achieves an average 147.25\% improvement over all instances, in contrast to 97.83\% and 102.94\% yielded by $FIP_{sg}$ and $FIP_{mt}$, respectively.

Moreover, in instances $MD_1$-$8$-$5$ and $SF_1$-$8$-$5$, the $BF$ and $FIP_{mt}$ obtained similar results in the parcel service criterion. Considering the increase in revenue, the $BF$ performs better overall, with values of 75.35\% and 47.98\% increase in comparison to 69.92\% and 45.62\% increase obtained by $FIP_{mt}$, for $MD_1$-$8$-$5$ and $SF_1$-$8$-$5$, respectively. This points to the fact that even though both methods effectively service the same proportion of parcels, considering all parcel and passenger requests simultaneously ultimately produces lower-cost routes. For instances $MD_2$-$30$-$20$, however, both FIP versions outperform the BF, even though the latter produced solutions with a greater parcel service rate. This is justified by the higher flexibility in parcel delivery of the FIP methods, which produces routes with lower travelling costs. 

It is worth noting that the $BF$ is more complex and time-consuming to solve. It could only find feasible solutions within the established time limit in 30\% and 3\% of instances in $MD_2$ and $SF_2$, respectively. The overall average gaps for these instance groups were respectively 4.72\% and 0.14\%.

\subsection{Computational time}

We further compare the $BF$ and both FIP variations in terms of computational time. Tables \ref{tab:smalltimes} and \ref{tab:medtimes} respectively present solution times for the three methods for groups of instances of classes $1$ and $2$.

Columns $n$ and $m$ represent each instance group's respective number of customer and parcel requests. The remainder of each table is divided into blocks of three columns, each representing a solution method for each dataset. The values in the table are the solution times, in seconds, averaged between the instances in each group.

\begin{table}[htbp]
\centering
\setlength{\tabcolsep}{0.15cm}
\small
\caption{Solution times (s) for $MD_1$, $GH_1$ and $SF_1$ datasets}
\begin{tabular}{ccccccccccccc}
\hline
 \multirow{3}{*}{\bf $n$} & \multirow{3}{*}{\bf $m$} & \multicolumn{3}{c}{\bf $MD_1$} & & \multicolumn{3}{c}{\bf $GH_1$} & & \multicolumn{3}{c}{\bf $SF_1$}\\\cline{3-5} \cline{7-9}\cline{11-13}
  &   & $BF$ & $FIP_{sg}$ & $FIP_{mt}$ & & $BF$ & $FIP_{sg}$ & $FIP_{mt}$ & & $BF$ & $FIP_{sg}$ & $FIP_{mt}$ \\\hline
5 & 5 & 0.1133 & 0.0017 & 0.0050 &  & 0.0940 & 0.0050 & 0.0050 &  & 0.0120 & 0.0100 & 0.0080 \\
5 & 6 & 0.1333 & 0.0050 & 0.0083 &  & 0.1410 & 0.0070 & 0.0080 &  & 0.0380 & 0.0060 & 0.0100 \\
5 & 7 & 0.2200 & 0.0100 & 0.0067 &  & 0.2250 & 0.0050 & 0.0080 &  & 0.0300 & 0.0120 & 0.0080 \\
5 & 8 & 0.2600 & 0.0117 & 0.0100 &  & 0.2530 & 0.0060 & 0.0130 &  & 0.0440 & 0.0060 & 0.0080 \\
5 & 9 & 0.4450 & 0.0083 & 0.0067 &  & 0.4490 & 0.0120 & 0.0100 &  & 0.0380 & 0.0060 & 0.0080 \\
5 & 10 & 0.4117 & 0.0117 & 0.0150 &  & 0.6410 & 0.0100 & 0.0080 &  & 0.0400 & 0.0180 & 0.0120 \\\hline
6 & 5 & 0.1867 & 0.0083 & 0.0050 &  & 0.2090 & 0.0050 & 0.0040 &  & 0.0300 & 0.0080 & 0.0120 \\
6 & 6 & 0.4000 & 0.0033 & 0.0100 &  & 0.2950 & 0.0070 & 0.0090 &  & 0.0600 & 0.0080 & 0.0040 \\
6 & 7 & 0.4617 & 0.0083 & 0.0083 &  & 0.3510 & 0.0100 & 0.0070 &  & 0.1160 & 0.0160 & 0.0060 \\
6 & 8 & 0.4267 & 0.0067 & 0.0117 &  & 0.5820 & 0.0110 & 0.0080 &  & 0.0600 & 0.0140 & 0.0100 \\
6 & 9 & 1.0117 & 0.0083 & 0.0100 &  & 0.9680 & 0.0120 & 0.0080 &  & 0.0620 & 0.0200 & 0.0140 \\\hline
7 & 5 & 0.4433 & 0.0017 & 0.0033 &  & 0.4510 & 0.0090 & 0.0110 &  & 0.0480 & 0.0140 & 0.0060 \\
7 & 6 & 0.7017 & 0.0117 & 0.0083 &  & 0.5120 & 0.0080 & 0.0120 &  & 0.0440 & 0.0080 & 0.0000 \\
7 & 7 & 0.7367 & 0.0117 & 0.0117 &  & 1.1870 & 0.0110 & 0.0100 &  & 0.0720 & 0.0120 & 0.0100 \\
7 & 8 & 1.1317 & 0.0033 & 0.0067 &  & 1.0980 & 0.0080 & 0.0090 &  & 0.0720 & 0.0140 & 0.0100 \\\hline
8 & 5 & 0.5817 & 0.0100 & 0.0067 &  & 0.7250 & 0.0100 & 0.0080 &  & 0.1000 & 0.0120 & 0.0060 \\
8 & 6 & 1.0350 & 0.0167 & 0.0150 &  & 1.1870 & 0.0060 & 0.0080 &  & 0.0880 & 0.0140 & 0.0160 \\
8 & 7 & 1.1650 & 0.0133 & 0.0117 &  & 1.4210 & 0.0160 & 0.0120 &  & 0.0660 & 0.0140 & 0.0080 \\ \hline
9 & 5 & 0.9733 & 0.0117 & 0.0100 &  & 1.2250 & 0.0110 & 0.0080 &  & 0.1100 & 0.0120 & 0.0140 \\
9 & 6 & 1.7867 & 0.0117 & 0.0100 &  & 2.0780 & 0.0140 & 0.0110 &  & 0.1960 & 0.0160 & 0.0080 \\\hline
10 & 5 & 1.4683 & 0.0083 & 0.0050 &  & 2.0780 & 0.0110 & 0.0070 &  & 0.1780 & 0.0180 & 0.0160 \\\hline
\end{tabular}
\label{tab:smalltimes}
\end{table}

\begin{table}[htbp]
\centering
\setlength{\tabcolsep}{0.2cm}
\small
\caption{Solution times (s) for $MD_2$ and $SF_2$ datasets}
\begin{tabular}{ccccccccc}
\hline
  \multirow{3}{*}{\bf $n$} & \multirow{3}{*}{\bf $m$} & \multicolumn{3}{c}{\bf $MD_2$} & & \multicolumn{3}{c}{\bf $SF_2$}\\\cline{3-5} \cline{7-9}
  &   & $BF$ & $FIP_{sg}$ & $FIP_{mt}$ & & $BF$ & $FIP_{sg}$ & $FIP_{mt}$ \\\hline
5 & 15 & 1.692 & 0.017 & 0.013 &  & 0.078 & 0.016 & 0.020 \\
5 & 20 & 3.168 & 0.017 & 0.018 &  & 0.512 & 0.032 & 0.022 \\
5 & 30 & 7.663 & 0.035 & 0.037 &  & 0.240 & 0.026 & 0.024 \\\hline
10 & 10 & 10.105 & 0.025 & 0.020 &  & 0.426 & 0.032 & 0.030 \\
10 & 15 & 29.275 & 0.025 & 0.018 &  & 0.742 & 0.050 & 0.048 \\
10 & 20 & 59.560 & 0.040 & 0.028 &  & 1.916 & 0.050 & 0.038 \\
10 & 30 & 562.842 & 0.062 & 0.048 &  & 3.102 & 0.082 & 0.146 \\\hline
15 & 10 & 49.617 & 0.023 & 0.018 &  & 2.692 & 0.030 & 0.028 \\
15 & 15 & 112.858 & 0.050 & 0.037 &  & 6.474 & 0.070 & 0.050 \\
15 & 20 & 2750.727 & 0.058 & 0.055 &  & 20.140 & 0.128 & 0.106 \\
15 & 30 & 3293.290 & 0.083 & 0.077 &  & 1476.108 & 0.128 & 0.128 \\\hline
20 & 10 & 285.185 & 0.048 & 0.038 &  & 11.588 & 0.086 & 0.076 \\
20 & 15 & 4964.738 & 0.053 & 0.047 &  & 48.786 & 0.122 & 0.064 \\
20 & 20 & 6275.758 & 0.060 & 0.048 &  & 49.034 & 0.152 & 0.254 \\
20 & 30 & 5865.410 & 0.055 & 0.050 &  & 87.970 & 0.206 & 0.140 \\\hline
30 & 10 & 6078.647 & 0.062 & 0.053 &  & 27.648 & 0.106 & 0.076 \\
30 & 15 & 7193.855 & 0.087 & 0.082 &  & 1530.634 & 0.236 & 0.218 \\
30 & 20 & 7193.098 & 0.134 & 0.118 &  & 475.478 & 0.338 & 0.410 \\
30 & 30 & -- & -- & -- &  & 3744.466 & 0.326 & 0.514 \\ \hline
\end{tabular}
\label{tab:medtimes}
\end{table}

Over all instances in class $1$, we can see the $BF$ performs well, solving the SARP optimally in 0.5 seconds, on average, with worst case of 2.08 seconds. Both FIP versions perform similarly in terms of computational times, reaching optimal solutions in 0.01 and 0.009 seconds on average and worst cases of 0.02 and 0.016 seconds for $FIP_{sg}$ and $FIP_{mt}$, respectively.

For instances of class $2$, we can observe that our method's performance deteriorates in terms of solution times, with average times of 24 minutes and worst case of two hours (the established time limit). Both versions of FIP produce solutions in under one second, with average values of 0.08 seconds for both and worst cases of 0.34 and 0.51 seconds for $FIP_{sg}$ and $FIP_{mt}$, respectively. 

The $BF$ performance in this criterion reinforces the difficulties associated with solving the SARP in a single stage, even with the use of request combination techniques.

\subsection{Efficiency of vehicle usage}\label{sec:effvehicle}

We extend our experiments to gauge vehicle usage efficiency when comparing the Bundle and FIP formulations. One of the primary motivations for adjoining parcel services to customer transportation systems is to make use of otherwise idle, relocating trips, to pick up and deliver parcels. Therefore, we decided to measure the deadheading distance obtained by the different strategies considered. 

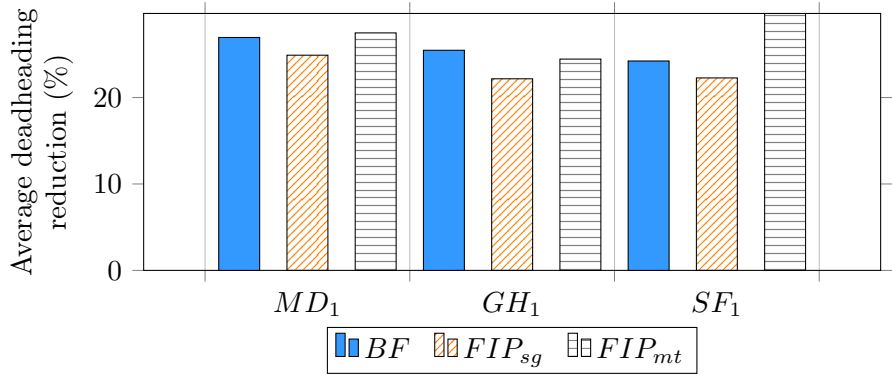
\begin{figure}[!htbp]
\begin{center}
\begin{tikzpicture}
\begin{axis}[
    legend columns=3, 
	xbar,
	width=0.7\textwidth, height=5cm, enlarge y limits=0,
        ylabel style={align=center},
	ylabel= Average deadheading\\reduction (\%),
	legend style={at={(0.5,-0.22)},anchor=north}, {/tikz/column 2/.style={column sep=7pt}},  {/tikz/column 4/.style={column sep=7pt}},
	symbolic x coords = {$MD_1$, $GH_1$, $SF_1$, K},
	ybar interval=0.6,
]
\addplot[draw=black, fill = color10]
	coordinates {($MD_1$,26.95) ($GH_1$,25.47)
		 ($SF_1$,24.23) (K, 0)};

\addplot[draw=black,  pattern=north east lines, pattern color=color9]
	coordinates {($MD_1$,24.90) ($GH_1$,22.17) 
		($SF_1$,22.27) (K, 0)};

\addplot[draw=black, pattern color=gray, pattern=horizontal lines]
	coordinates {($MD_1$,27.48) ($GH_1$,24.45) 
		($SF_1$,29.73)(K, 0)};

\legend{$BF$,$FIP_{sg}$,$FIP_{mt}$}
\end{axis}
\end{tikzpicture}
\captionof{figure}{Average deadheading reduction (\%) for $MD_1$, $GH_1$ and $SF_1$ datasets}
\label{fig:idles}
\end{center}
\end{figure}

\begin{figure}[!htbp]
\begin{center}
\begin{tikzpicture}
\begin{axis}[
    legend columns=3, 
	xbar,
	width=0.7\textwidth, height=5cm, enlarge y limits=0,
        ylabel style={align=center},
	ylabel= Average deadheading\\reduction (\%),
	legend style={at={(0.5,-0.22)},anchor=north}, {/tikz/column 2/.style={column sep=7pt}},  {/tikz/column 4/.style={column sep=7pt}},
	symbolic x coords = {$MD_2$, $SF_2$, K},
	ybar interval=0.6,
]
\addplot[draw=black, fill = color10]
	coordinates {($MD_2$,28.01) ($SF_2$,25.08) (K, 0)};

\addplot[draw=black,  pattern=north east lines, pattern color=color9]
	coordinates {($MD_2$,21.44) ($SF_2$,22.10) (K, 0)};

\addplot[draw=black, pattern color=gray, pattern=horizontal lines]
	coordinates {($MD_2$,25.69) ($SF_2$,31.30) (K, 0)};

\legend{$BF$,$FIP_{sg}$,$FIP_{mt}$}
\end{axis}
\end{tikzpicture}
\captionof{figure}{Average deadheading reduction (\%) for $MD_2$ and $SF_2$ datasets}
\label{fig:idlem}
\end{center}
\end{figure}
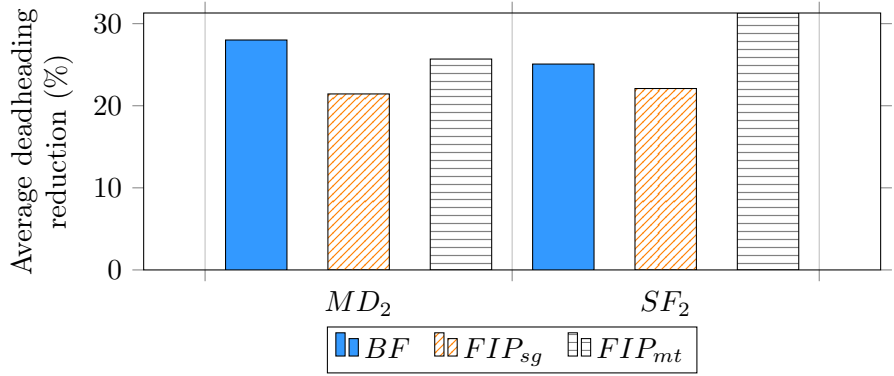

We calculate the decrease in the percentage of deadheading distance between a passenger-only scenario the FIP and BF. The results are averaged between all instances of each dataset and separated according to their respective classes, presented in Figures \ref{fig:idles} and \ref{fig:idlem}.


In all methods, there is some reduction in deadheading percentage when compared to a passenger-only solution. This reinforces the notion that the joint transportation of parcels and people reduces the overall number of relocating trips and unnecessary $CO_2$ emissions.

Both the $BF$ and $FIP_{mt}$ yield better results than $FIP_{sg}$ in this criterion. For instances in class $1$, on average, there is a reduction of 25.55\%, 23.11\%, and 27.22\%, respectively, for the BF, $FIP_{sg}$ and $FIP_{mt}$. For instances in class $2$, we can see an average reduction of 26.55\%, 21.77\%, and 28.49\%, respectively, for the BF, $FIP_{sg}$ and $FIP_{mt}$, as well as a larger difference between the values of $FIP_{sg}$ and both the $BF$ and $FIP_{mt}$, when compared to the results for instances 
of class $1$.

Both FIP variations benefit from the fact that parcel deliveries can take place at any point in the route; however, there is a further advantage of $FIP_{mt}$ in comparison to $FIP_{sg}$ due to limitations in parcel carrying capacity from the latter.
The $BF$ has the same parcel capacity limitations as $FIP_{sg}$, and, furthermore, the parcel deliveries must occur after the service of exactly one passenger. Nevertheless, our proposed method produces competitive results, surpassing $FIP_{mt}$, on average, in two of the five datasets considered, namely $GH_1$ and $MD_2$. This confirms that solving the SARP and producing routes considering both passenger and parcel requests simultaneously leads to better quality solutions in terms of deadheading reduction criterion, even for instances of larger size, as in $MD_2$.

\section{Concluding remarks and future work}

In this work, we have considered the problem of people and parcels sharing trips in ride-hailing systems and propose a novel MILP formulation to a version of SARP, a problem introduced by \cite{Li2014}. This novel formulation tackles the problem by combining customer and parcel requests in bundles and using strategies inspired by the GVRP to build routes with them. We compared the proposed formulation to two versions of an existing model for this problem, namely $FIP_{sg}$ and $FIP_{mt}$, which are part of a two-stage strategy for the SARP.

We used four criteria to compare the solution methods: the amount of serviced parcels, the increase in revenue from a passenger-only solution, the solving times, and the efficiency of vehicle usage.

Our findings showed that our $BF$ performs better than the FIP methods in both financial and parcel service criteria. Our method was able to service more parcels in 98\% and 100\% of the instance groups tested, compared to $FIP_{mt}$ and $FIP_{sg}$, respectively. Furthermore, in 99\% of instance groups, the $BF$ produced better quality solutions, achieving higher revenues than the other considered methods. Even under more constrained assumptions regarding parcel service, the $BF$ leads to higher deadheading reduction against $FIP_{mt}$, in two out of five instance sets and in all five instance sets when compared to $FIP_{sg}$. These results point to the visible benefits of tackling the SARP in a single stage, as opposed to attempts of parcel insertion in a pre-defined passenger solution.

The research avenues arising from this work are numerous. 
First, they can focus on solution methods for more complex versions of SARP, including additional flexibility on the rules regarding parcel delivery and vehicle capacity. Additionally, we primarily focused on exact solution approaches to achieve an unbiased (i.e., optimal) assessment of the economic potential of passenger and freight transport integration. Since the computational effort needed to solve the $BF$ still remains a bottleneck for larger instances, we also recommend progressing on heuristic approaches capable of solving integrated routing and freight insertion problems at scale. With this vision, the exact solutions will also prove extremely useful for future benchmarking and solution-quality assessment.

\section*{Acknowledgements}

This research was partially supported by the Brazilian research agencies CNPq, grants 309580/2021-8, 406245/2021-5, 307843/2018 and 308528/2018-2, FAPERJ, grant E-26/202.790/2019, and FAPESQ, grant 261/2020.
\clearpage
\bibliographystyle{apacite-no-initials}
\bibliography{references}

\end{document}